\documentclass[journal]{IEEEtran}
\usepackage{amsmath,amsfonts,amssymb}
\usepackage{amsthm}
\usepackage{algorithm}
\usepackage{algorithmic}
\usepackage{array}
\usepackage{enumitem}
\usepackage{acronym}
\usepackage[caption=false,font=footnotesize]{subfig}
\usepackage{textcomp}
\usepackage{stfloats}
\usepackage{booktabs}
\usepackage{multirow}
\usepackage{url}
\usepackage{graphicx}
\usepackage{tikz}
\IfFileExists{dsfont.sty}{\usepackage{dsfont}}{\newcommand{\mathds}[1]{\mathbf{##1}}}
\usepackage{pifont}
\usepackage{balance}
\makeatletter
\newcommand\blfootnote[1]{%
  \begingroup
  \renewcommand\thefootnote{}\footnotetext{#1}%
  \addtocounter{footnote}{-1}%
  \endgroup
}
\makeatother
\allowdisplaybreaks[4]
\AtBeginDocument{\setlength{\abovedisplayskip}{3pt plus 1pt minus 1pt}\setlength{\belowdisplayskip}{3pt plus 1pt minus 1pt}\setlength{\abovedisplayshortskip}{1pt plus 1pt}\setlength{\belowdisplayshortskip}{2pt plus 1pt minus 1pt}}
\newtheorem{theorem}{Theorem}
\newtheorem{proposition}{Proposition}
\newtheorem{lemma}{Lemma}
\newcommand{\cmark}{\ding{51}}
\newcommand{\xmark}{\ding{55}}

\newcommand{\sinr}{\mathrm{SINR}}
\newcommand{\TN}{\mathrm{TN}}
\newcommand{\NTN}{\mathrm{NTN}}

\input{Acronyms}

\begin{document}
\bstctlcite{IEEEexample:BSTcontrol}

\title{SINR-Aware Adaptive Offloading and Resource Allocation in Shared-Spectrum Integrated Terrestrial and Non-Terrestrial Networks}
\author{Aman~Khalid, Muhammad~Sohaib~Aslam, Ahmed~Naeem, Seyit~Karatepe, Muhammet~Kirik, Hüseyin~Arslan,~\IEEEmembership{Fellow,~IEEE}
\thanks{A. Khalid, M. S. Aslam, A. Naeem are with the School of Electrical Engineering and Computer Science, National University of Sciences and Technology, Islamabad 44000, Pakistan (email: amkhalid.bee23seecs@seecs.edu.pk, maslam.bee23seecs@seecs.edu.pk, ahmed.naeem@seecs.edu.pk). \\
S. Karatepe is with Türk Telekom R\&D, Ankara, 06080, Türkiye (email: seyit.karatepe@turktelekom.com.tr).\\
M. Kirik and H. Arslan are with the Department of Electrical and Electronics
Engineering, Istanbul Medipol University, Istanbul, 34810, Türkiye (email: muhammet.kirik@medipol.edu.tr, huseyinarslan@medipol.edu.tr).}\\This work has been submitted to the IEEE for possible publication. Copyright may be transferred without notice, after which this version may no longer be accessible.}
\markboth{IEEE Transactions on Communications}%
{Khalid \MakeLowercase{\textit{et al.}}: SINR-Aware Adaptive Offloading and Resource Allocation in Shared-Spectrum Integrated Terrestrial and Non-Terrestrial Networks}

\maketitle
\begin{abstract}
Sharing the sub-6~GHz spectrum between terrestrial networks and low Earth orbit satellites introduces cross-tier interference in both directions, as terrestrial transmissions can degrade satellite-link reception, while satellite transmissions can impair the performance of terrestrial links. Conventional reference signal received power (RSRP)-based offloading, designed for separate frequency bands, does not account for this cross-tier interference. We study offloading and resource allocation in an integrated terrestrial and non-terrestrial network through a two-timescale framework in which signal-to-interference-plus-noise ratio (SINR)-aware adaptive traffic offloading (SAATO) updates user associations from filtered SINR and load-aware expected throughput, switching a user only across a hysteresis band, while joint cross-tier fractional programming (Joint-FP) optimizes resource allocation for the resulting associations. Under an equal-share resource allocation model, the resulting decision metric approximates the change in proportional-fair utility with an error that decays with load. We prove that joint allocation under the sum-rate utility is nondeterministic polynomial-time (NP)-hard and that the utility sequence accepted by Joint-FP is non-decreasing and convergent. At full spectral overlap, RSRP-based offloading leaves 44\% of user samples below 0.5~Mbps (outage), whereas SAATO with Joint-FP, given instantaneous channel knowledge, leaves none and increases geometric-mean throughput by 8\% and 2\% under half and full spectral overlap, respectively, relative to power control without offloading. It also limits tier switching to at most 0.05 switches per user per minute.
\end{abstract}

\acresetall

\section{Introduction}
\label{sec:introduction}

\IEEEPARstart{T}{errestrial} networks alone cannot provide both the capacity and the geographic coverage that \ac{6G} is expected to demand \cite{saad2020vision, giordani2021ntn6g}. Traffic keeps growing, and emerging services need higher data rates and lower latency than \ac{5G} was designed to support \cite{saad2020vision}. Extending ground infrastructure to rural and hard-to-reach areas, or to disaster zones, is too costly \cite{kodheli2021satcom, giordani2021ntn6g}. \Acp{NTN}, such as \ac{LEO} satellites, can serve areas that ground infrastructure does not reach and can also carry part of the traffic of congested cells. Their integration with \acp{TN} yields an \ac{ITNTN} \cite{rahman2025joint}, which is expected to be an integral part of \ac{6G}. Direct service to handsets is moving into practice: \ac{3GPP} specified \ac{NR} support for \acp{NTN} in Release~17 \cite{3gpp38300} and Release~18 \cite{saad2024ntn}, and the U.S. \ac{FCC} now allows supplemental satellite coverage in terrestrial mobile spectrum \cite{fcc2024scs}.

The benefit of integrating the two tiers depends on how they use the spectrum. Spectrum is scarce, which leads the tiers of an \ac{ITNTN} to share the same band and to cause cross-tier interference that degrades network performance. Within the terrestrial tier, cell-edge users suffer most from a weak desired signal and strong interference from neighboring \acp{gNB}. With separate bands, an offloaded cell-edge user would escape the cross-tier interference and release terrestrial resources, but a shared band offers no separate spectrum to escape. There, interference flows in both directions: the \acp{gNB} interfere with the satellite link of users inside the cells, and the satellite in turn can add interference at terrestrial receivers \cite{lee2021reverse, lee2023feasibility, kim2024leotnsharing}. Moving a cell-edge user to the satellite does not necessarily improve its \ac{SINR}, because nearby \acp{gNB} can remain dominant on the shared \acp{PRB}. An offloading rule must therefore judge each user by the \ac{SINR} it would obtain on the satellite link, not by the weakness of its terrestrial link.

The motion of a \ac{LEO} satellite complicates the selection of users to offload. At an altitude of 600~km \cite{3gpp38821}, the satellite stays above $30^\circ$ elevation for only about four minutes per pass by orbital geometry, and the \ac{LOS} condition and desired signal of each satellite user change with the elevation angle. A rule that follows every new \ac{SINR} measurement would move users back and forth between the tiers, and each switch requires \ac{RRC} signaling \cite{3gpp38331}. Offloading must weigh the \ac{SINR} of both tiers, the time variation of the satellite link and the cost of each switch.

\subsection{Related Works}
\label{subsec:related_work}
Offloading terrestrial traffic to a satellite is a familiar remedy for congested cells, and most work on it gives the satellite a carrier of its own. In \cite{abderrahim2022proactive}, small base stations predict their traffic and hand the excess to a multi-satellite network, whereas in \cite{nguyenkha2024} \ac{LEO} satellites provide backhaul to isolated base stations and user and base-station association are optimized jointly. With the satellite on its own spectrum, interference between the tiers does not enter the offloading decision. The problem changes when both tiers share a band, because moving a user to the satellite then affects not only the load on each tier but also the interference seen by users on the same resources. The closest work to ours \cite{rahman2025joint} offloads inside an \ac{ITNTN} with a geostationary satellite: it offloads a user that has a \ac{LOS} satellite link when its \ac{RSRP} is low or when it sits at the cell edge, and it allocates power separately in each tier. Neither criterion considers the \ac{SINR} that the user would obtain after switching, and that \ac{SINR} can be very low when nearby \acp{gNB} transmit on the same \acp{PRB}. Max-\ac{SINR} association \cite{andrews2014loadbalancing} does use link quality, but it ignores tier loads and does not damp tier switches.

Association is only part of the coexistence problem, because each tier's performance in a shared band depends on the resources and powers of the other. The \ac{3GPP} study for \acp{NTN} covers adjacent carriers only \cite{3gpp38863}, and a co-channel study examines how satellite transmissions affect the terrestrial network, finding little impact in its scenario \cite{lee2023feasibility}. On the spectrum side, reverse pairing lets the satellite \ac{UL} reuse the terrestrial \ac{DL} band \cite{lee2021reverse, lee2023feasibility}, although unmodified handsets cannot receive in the \ac{UL} band, and a stochastic-geometry analysis finds that the preferred pairing depends on the ratio of base-station to user density \cite{kim2024leotnsharing}. On the power side, cognitive satellite--terrestrial networks adapt the power of a secondary tier to protect the primary receivers \cite{vassaki2013cognitive}, \ac{QoS}-aware power control with spatial nulling is more recent \cite{tsampazi2026fr3}, and \cite{heydarishahreza2024spectrum} surveys the wider options. These works manage spectrum use or interference, but none of them decides which tier serves each user.

Even with a suitable association metric, users should not switch tiers whenever their measured link conditions fluctuate. Across a satellite cell the \ac{RSRP} varies by only a few decibels, which forces handover triggers to trade radio link failures against ping-pong handovers \cite{juan2022leohandover}. A tier switch faces the same trade-off. Power-, elevation- and distance-based triggers have been compared for \ac{LEO} links \cite{kodheli2024handover}, and conditional handover, which switches only after a preconfigured condition has held, has been evaluated there \cite{juan2022cho} at the price of extra handovers. These studies address mobility-related switching, whereas \ac{SAATO} carries the same idea of switching only after a condition has held over to tier selection, using a filtered metric and a hysteresis band.

A stable association fixes which link serves each user, but not how the two tiers should share power on the same resources. Because each rate depends on a \ac{SINR}, a ratio of powers, \ac{FP} is the natural tool. The quadratic transform \cite{shen2018fp} replaces each ratio with a term in an auxiliary variable, and the Lagrangian dual transform \cite{shen2018fp2} moves the ratios out of the logarithms of a weighted sum rate; both converge to a stationary point. \ac{FP} has allocated \ac{DL} power for a geostationary satellite \cite{rahman2023milcom} and for the orthogonal-spectrum \ac{ITNTN} \cite{rahman2025joint}, but, to our knowledge, not for two tiers with separate power budgets whose transmissions fall on one resource grid.

The remaining challenge is to bring these decisions together. In a shared-band \ac{ITNTN}, the serving tier must be chosen with the \ac{SINR} and load of both tiers in view, the power allocation must charge each transmitter for the interference it causes in the other tier, and tier switches must be damped. Unlike cognitive formulations, which protect a primary receiver, our setting serves users on both tiers and treats the tier of each user as a decision. \ac{SAATO} addresses the first and third requirements and \ac{Joint-FP} the second, and Table~\ref{tab:comparison} shows where the closest works fall short.

\begin{table}[!t]
\centering
\caption{Comparison With Related Work}
\label{tab:comparison}
\footnotesize
\renewcommand{\arraystretch}{1.0}
\setlength{\tabcolsep}{3pt}
\begin{tabular}{|c|c|c|c|c|c|}
\hline
\textbf{Ref.} & \textbf{Shared} & \textbf{Tier} & \textbf{Cross-tier} & \textbf{Cross-tier} & \textbf{Load-aware,} \\
 & \textbf{band} & \textbf{selection} & \textbf{interf.} & \textbf{power} & \textbf{damped} \\
\hline
\cite{rahman2025joint} & \xmark & \cmark & \xmark & \xmark & \xmark \\ \hline
\cite{lee2021reverse} & \cmark & \xmark & \cmark & \xmark & \xmark \\ \hline
\cite{kim2024leotnsharing} & \cmark & \xmark & \cmark & \xmark & \xmark \\ \hline
\cite{vassaki2013cognitive} & \cmark & \xmark & \cmark & \cmark & \xmark \\ \hline
\cite{tsampazi2026fr3} & \cmark & \xmark & \cmark & \cmark & \xmark \\ \hline
This work & \cmark & \cmark & \cmark & \cmark & \cmark \\ \hline
\end{tabular}
\end{table}
\blfootnote{\textit{Notation:} Bold symbols ($\mathbf{P}$, $\boldsymbol{V}$, $\boldsymbol{E}$, $\mathbf{x}$, $\mathbf{y}$) collect scalar variables over users and \acp{PRB}. Calligraphic letters denote sets, except for $\mathcal{P}_0$, the circularly symmetric complex Gaussian distribution $\mathcal{CN}(0,1)$ and the complexity order $\mathcal{O}(\cdot)$. The operator $|\cdot|$ denotes cardinality or magnitude. The superscript $m$ indexes \acp{PRB}, and the pairs $(k,l)$ and $(i,j)$ denote a \ac{gNB} with one of its users and a beam with one of its users. The operators $\lfloor\cdot\rceil$ and $\lceil\cdot\rceil$ denote rounding to the nearest integer (ties rounded up) and the ceiling, $\mathds{1}\{\cdot\}$ is the indicator function and $\exp_2(z)=2^z$.}
\subsection{Contributions and Paper Organization}
\label{subsec:contributions}
Existing studies on shared-spectrum \acp{ITNTN} do not fully address the coupled challenges of interference-aware traffic offloading, joint cross-tier resource allocation and tier-switching stability. This paper addresses these gaps through the following contributions:

\begin{itemize}[leftmargin=*]
\item \textbf{Shared-spectrum interference modeling:} We formulate a coupled \ac{SINR} model with adjustable spectral overlap, accounting for intra-tier and cross-tier interference in a shared-band \ac{ITNTN} comprising a \ac{LEO} satellite and a terrestrial network.

\item \textbf{\ac{SINR}-aware adaptive offloading:} The proposed \ac{SAATO} scheme selects the serving tier of each user by comparing the expected throughput derived from the filtered \acp{SINR} and tier loads, rather than relying on \ac{RSRP}. A hysteresis band prevents unnecessary switching between tiers, while an equal-share analysis shows that the offloading decision metric approximates the change in the network \ac{PF} utility under equal resource sharing.

\item \textbf{Joint cross-tier resource allocation:} We develop \ac{Joint-FP}, which optimizes transmit power and \ac{PRB} allocation across both tiers, accounting for cross-tier interference through interference leakage terms. We prove that the sum-rate maximization problem is \ac{NP}-hard and establish that the utility values accepted by \ac{Joint-FP} form a non-decreasing, convergent sequence.

\item \textbf{System-level performance evaluation:} We evaluate the proposed framework under different spectral-overlap levels and channel uncertainties. At full overlap, \ac{RSRP}-based offloading leaves 44\% of user samples in outage, whereas \ac{SAATO} with \ac{Joint-FP} leaves no user sample below 0.5~Mbps given instantaneous channel knowledge. Relative to power control without offloading, the proposed framework improves geometric-mean throughput by 8\% and 2\% at half and full overlap, respectively, while limiting tier switching to at most 0.05 switches per user per minute.
\end{itemize}

\par The rest of the paper is organized as follows. Section~\ref{sec:system_model} presents the system model, Section~\ref{sec:proposed_framework} the \ac{SAATO} scheme and Section~\ref{sec:solution} \ac{Joint-FP}, and Section~\ref{sec:simulation} reports the simulation results. Section~\ref{sec:conclusion} concludes the paper.

\section{System Model and Coupled Cross-Tier SINR Formulation}
\label{sec:system_model}
Offloading and power control can only be judged in a model where the two tiers genuinely interfere. We describe the shared-band network and its assumptions, the channels, the coupled \ac{SINR} expressions and, finally, the optimization problem that the rest of the paper solves.

\subsection{Shared-Spectrum Architecture and Coexistence Topology}
\label{subsec:architecture}

\begin{figure*}[!t]
\centering 
\resizebox{\textwidth}{!}{
\includegraphics{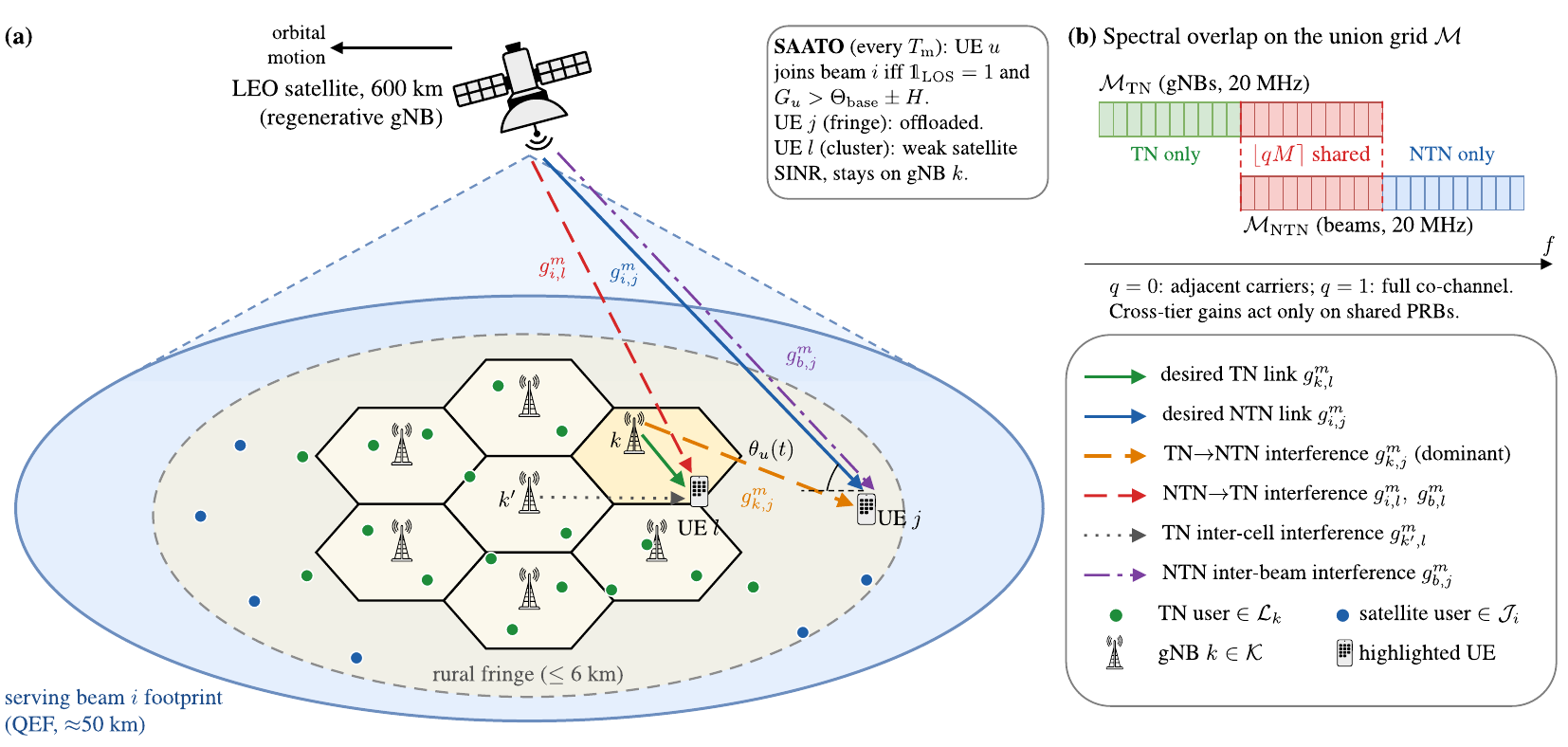}}
\caption{System model: (a) \acs{LEO} satellite and a seven-cell terrestrial cluster; (b) the two carriers share $\lfloor q M\rceil$ \acsp{PRB}.}
\label{fig:system}
\end{figure*}

The network has $K$ terrestrial macro \acp{gNB}, indexed by $\mathcal{K}=\{1,\dots,K\}$, in a hexagonal cluster of cell radius $R_{\mathrm{cell}}$, and one \ac{LEO} satellite acting as a regenerative \ac{gNB} \cite{3gpp38821}. The satellite forms $I$ \ac{QEF} beams, indexed by $\mathcal{I}=\{1,\dots,I\}$, with boresights fixed on ground spots as it moves. With a \ac{FRF} $N_{\mathrm{re}}=3$, we model the nearest co-channel beams $b\in\mathcal{B}$ as full-load interferers on every \ac{PRB} of the satellite carrier. The offloading decision assigns each user either to the set $\mathcal{L}_k$ of terrestrial users $l$ of \ac{gNB} $k$ or to the set $\mathcal{J}_i$ of satellite users $j$ of beam $i$. The $N_{\mathrm{u}}$ users, indexed by $u$, include rural fringe users beyond the cell edges (Fig.~\ref{fig:system}(a)).

Both tiers use \ac{NR} carriers in the 2~GHz S-band reference scenario of \cite{3gpp38821}, with $f_c=2$~GHz, a bandwidth $B=20$~MHz and a \ac{SCS} $\Delta f=30$~kHz. Each carrier has $M=51$ \acp{PRB}, and we schedule at \ac{PRB} granularity. We place the satellite carrier on a common grid such that, for an overlap fraction $q$, the \ac{PRB} sets $\mathcal{M}_{\TN}$ and $\mathcal{M}_{\NTN}$ satisfy
\begin{equation}
\big|\mathcal{M}_{\TN}\cap\mathcal{M}_{\NTN}\big| = \lfloor q M \rceil, \qquad q\in[0,1].
\label{eq:overlap}
\end{equation}
The union grid $\mathcal{M}=\mathcal{M}_{\TN}\cup\mathcal{M}_{\NTN}$ then contains $M_{\cup}=2M-\lfloor q M\rceil$ \acp{PRB}. For $q=0$ the carriers are adjacent, as in \cite{3gpp38863}, while for $q=1$ they coincide. We neglect leakage into adjacent \acp{PRB}, which confines cross-tier interference to the overlapping \acp{PRB}, where it acts at full strength (Fig.~\ref{fig:system}(b)).

For the terrestrial tier, we adopt the multi-user multiple-input multiple-output (MU-MIMO)\acused{MU-MIMO}\acused{MIMO} model of \cite{rahman2025joint}, realized here with \ac{ZF} precoding, in which each \ac{gNB} co-schedules up to $\mu_{\TN}$ single-antenna users per \ac{PRB}. We refer to $\mu_{\TN}$ as the MU-MIMO order. Imperfect \ac{CSI} leaves a fraction $\xi_{\TN}$ of the co-scheduled power as residual intra-cell interference. A satellite beam covers an area much larger than the terrestrial cluster, gives no useful spatial separation and serves $\mu_{\NTN}=1$ user per \ac{PRB}. If a beam holds more users than \acp{PRB}, users sharing a \ac{PRB} interfere at full power ($\xi_{\NTN}=1$). We also assume:
\begin{enumerate}[label=\textbf{A\arabic*:}, leftmargin=*, itemsep=0pt]
\item \textbf{Association:} At any moment, a user is served by exactly one \ac{gNB} or one beam \cite{rahman2025joint}.
\item \textbf{Timing:} Each \ac{UE} pre-compensates delay and Doppler using its \ac{GNSS} position and broadcast ephemeris \cite{3gpp38821}. The service-link round trip stays below 8~ms above $30^\circ$ elevation. We update association every $T_{\mathrm{m}}$ and resources every $T_{\mathrm{e}}$ seconds.
\item \textbf{Power budgets:} \Ac{gNB} $k$ and beam $i$ are allotted budgets $P_{k,\max}$ and $P_{i,\max}$ \cite{rahman2025joint}. The TR~38.821 \ac{EIRP} density of 34~dBW/MHz gives 47~dBW over 20~MHz; the 30~dBi antenna gain \cite{3gpp38821} leaves $P_{i,\max}=17$~dBW (47~dBm).
\item \textbf{Terminals:} Satellite users have no angular protection against terrestrial signals, since every \ac{UE} is a handheld with an omnidirectional antenna \cite{3gpp38821}.
\item \textbf{Side information:} The \ac{LOS} states are known to the network. The satellite \ac{LOS} state changes only on a scale of tens of seconds (Section~\ref{subsec:channel_models}), far longer than $T_{\mathrm{m}}$; the network can therefore estimate it from the elevation, the Rician factor or the stability of the measured \ac{RSRP}. We treat it as known to isolate the offloading logic. Each \ac{UE} reports its strongest \ac{gNB} $k(u)$ by \ac{RSRP} and its covering beam $i(u)$. Intra-tier handover is not modeled, and all users start on the \ac{TN}. The \acp{SS-SINR} count all other co-channel transmitters at full load and omit intra-cell leakage.

\item \textbf{Scope:} Each allocation uses instantaneous per-\ac{PRB} channel knowledge, rates follow Shannon's formula, and one satellite serves a rural fringe. Multi-satellite operation, the \ac{UL} and delayed channel reports lie outside the model.
\end{enumerate}

\subsection{3GPP Channel Models}
\label{subsec:channel_models}

Both tiers follow 3GPP channel models.
\subsubsection{Terrestrial Channel Model}
The channel gain between \ac{gNB} $k$ and \ac{UE} $u$ on \ac{PRB} $m$ at time $t$ is
\begin{equation}
g_{k,u}^m(t) = G_{\mathrm{t}}\, G_{\mathrm{r}}\, 10^{-(L_{k,u}+X_{k,u})/10}\, \big|h_{k,u}^m(t)\big|^2\, \mathds{1}\{m\in\mathcal{M}_{\TN}\},
\label{eq:tn_channel_gain}
\end{equation}
where $G_{\mathrm{t}}$ and $G_{\mathrm{r}}$ are the \ac{gNB} and \ac{UE} antenna gains. The path loss $L_{k,u}$, \ac{LOS} probability and log-normal shadowing $X_{k,u}$ come from the TR~38.901 \ac{RMa} model \cite{3gpp38901}. Fast fading on \ac{LOS} links is Rician, $h_{k,u}^m(t) = \sqrt{K_{\mathrm{R},\TN}/(K_{\mathrm{R},\TN}+1)}\, e^{\mathrm{j}\,\mathrm{ph}_{k,u}} + \sqrt{1/(K_{\mathrm{R},\TN}+1)}\, h_{k,u}^{\mathrm{d},m}(t)$, where $K_{\mathrm{R},\TN}$ denotes the Rician factor, $\mathrm{j}=\sqrt{-1}$, $\mathrm{ph}_{k,u}$ is a phase uniform in $[0,2\pi)$ and $h_{k,u}^{\mathrm{d},m}(t)\sim\mathcal{CN}(0,1)$ is the diffuse component, whereas \ac{NLOS} links are Rayleigh ($K_{\mathrm{R},\TN}=0$). Fading is constant over blocks of $N_{\mathrm{coh}}$ adjacent \acp{PRB}, about one coherence bandwidth, and the terrestrial \ac{LOS} state and shadowing stay fixed during a pass.

\subsubsection{Satellite Channel Model}
The ground track of the satellite's circular orbit crosses the cluster at a random azimuth and offset. Over the cluster center, the peak elevation is uniform on $[\theta_{\mathrm{pk}},90^\circ]$, where $\theta_{\mathrm{pk}}$ is its lower bound, and service lasts while the elevation is at least $\theta_{\min}$. Let $d_u(t)$ and $\theta_u(t)$ denote the slant range and elevation seen by \ac{UE} $u$, and $\mathrm{ang}_{i,u}(t)$ its off-boresight angle in beam $i$. Beam $i$ reaches $u$ with a channel gain of
\begin{align}
g_{i,u}^m(t) ={}& G_{\mathrm{s}}\,G_{\mathrm{r}}\,A\big(\mathrm{ang}_{i,u}(t)\big)\,10^{-L^{\mathrm{sat}}_u(t)/10} \nonumber\\
&\times \big|h_u^{\mathrm{sat}}(t)\big|^2\,\mathds{1}\{m\in\mathcal{M}_{\NTN}\},
\label{eq:ntn_channel_gain}
\end{align}
and swapping $i$ for $b$ yields the gain $g_{b,u}^m$ of an interfering beam. The peak antenna gain is $G_{\mathrm{s}}$, and $A(\mathrm{ang})=|2J_1(c_{\mathrm{ap}}\sin\mathrm{ang})/(c_{\mathrm{ap}}\sin\mathrm{ang})|^2$ gives the circular-aperture pattern of TR~38.811 \cite{3gpp38811}. Here $J_1(\cdot)$ is the first-order Bessel function of the first kind, and $c_{\mathrm{ap}}$ is set such that the 3-dB beamwidth equals $\mathrm{ang}_{3\mathrm{dB}}$.

On a hexagonal beam lattice, the nearest co-channel beams lie $\sqrt{3N_{\mathrm{re}}}$ beam radii apart. With the beam radius set to half the 3-dB beamwidth, they are $\sqrt{3N_{\mathrm{re}}}\,\mathrm{ang}_{3\mathrm{dB}}/2$ apart, which gives a carrier-to-interference ratio of about 10~dB at the beam center. The satellite path loss in dB is $L^{\mathrm{sat}}_u = 32.45 + 20\log_{10} f_c + 20\log_{10} d_u + X^{\mathrm{sat}}_u + L^{\mathrm{CL}}_u + L_{\mathrm{pol}} + L_{\mathrm{sc}} + L_{\mathrm{atm}}$, that is, the free-space loss, with $f_c$ in GHz and $d_u$ in meters, plus the shadowing, clutter, polarization, scintillation and atmospheric terms defined next. The \ac{LOS} probability, shadowing $X^{\mathrm{sat}}_u$ and clutter loss $L^{\mathrm{CL}}_u$ (\ac{NLOS} only) take elevation-dependent rural S-band values from \cite{3gpp38811}. Each user draws one uniform variable per pass; its \ac{LOS} state changes only when the elevation moves the \ac{LOS} probability across that draw. Polarization mismatch $L_{\mathrm{pol}}$, scintillation $L_{\mathrm{sc}}$ and atmospheric absorption $L_{\mathrm{atm}}$ follow the TR~38.821 link budget \cite{3gpp38821}. Fast fading $h_u^{\mathrm{sat}}$ is frequency-flat given the small delay spread, Rayleigh in \ac{NLOS} and Rician in \ac{LOS}, with the Rician factor $K_{\mathrm{R},\NTN}$ near that of the \ac{NTN} \ac{TDL}-C profile of TR~38.811 \cite{3gpp38811}. The cross-tier gain $g_{i,l}^m$ from a beam to a terrestrial user follows the satellite model of \eqref{eq:ntn_channel_gain} and the path loss above, and $g_{k,j}^m$ from a \ac{gNB} to a satellite user follows the terrestrial model of \eqref{eq:tn_channel_gain} and its Rician fading.
\subsection{Coupled Downlink SINR and Achievable Rates}
\label{subsec:sinr_formulation}
The channels fix how strongly each transmitter reaches each receiver, and together they give the \ac{SINR} of every user.
We write $P_{k,l}^m$ and $P_{i,j}^m$ for the transmit powers, in mW, from \ac{gNB} $k$ to terrestrial user $l$ and from beam $i$ to satellite user $j$ on \ac{PRB} $m$. Binary variables $V_{k,l}^m$ and $E_{i,j}^m$ show whether $l$ and $j$ are scheduled on $m$, while every interfering beam, with the same budget $P_{i,\max}$ as the serving beam, radiates $\bar{P}^m=(P_{i,\max}/M)\,\mathds{1}\{m\in\mathcal{M}_{\NTN}\}$. For a terrestrial user $l\in\mathcal{L}_k$ on \ac{PRB} $m\in\mathcal{M}_{\TN}$, the \ac{SINR} is
\begin{equation}
\sinr_{k,l}^{\TN}(m) = \frac{V_{k,l}^m P_{k,l}^m g_{k,l}^m}{Z_{k,l}^m},
\label{eq:sinr_tn}
\end{equation}
where the interference-plus-noise power is
\begin{align}
Z_{k,l}^m ={}& \underbrace{\xi_{\TN}\!\sum_{l'\neq l}V_{k,l'}^m P_{k,l'}^m g_{k,l}^m}_{\text{intra-cell}}
+\underbrace{\sum_{k'\neq k}\sum_{l'\in\mathcal{L}_{k'}}\!V_{k',l'}^m P_{k',l'}^m g_{k',l}^m}_{\text{terrestrial inter-cell}} \nonumber\\
&+\underbrace{\sum_{i\in\mathcal{I}}\sum_{j\in\mathcal{J}_i}E_{i,j}^m P_{i,j}^m g_{i,l}^m+\sum_{b\in\mathcal{B}}\bar{P}^m g_{b,l}^m}_{\text{cross-tier NTN-to-TN}}+\sigma^2.
\label{eq:interf_tn}
\end{align}
Here, $\sigma^2=N_0B_{\mathrm{RB}}10^{N\!F/10}$ is the thermal noise power in one \ac{PRB} of bandwidth $B_{\mathrm{RB}}=12\Delta f$, with $N_0$ the noise density and $N\!F$ the noise figure. All quantities in the equations are linear: dB, dBi and dBm values are converted by $10^{(\cdot)/10}$, and powers are in mW. Likewise, for a satellite user $j\in\mathcal{J}_i$ on \ac{PRB} $m\in\mathcal{M}_{\NTN}$,
\begin{equation}
\sinr_{i,j}^{\NTN}(m) = \frac{E_{i,j}^m P_{i,j}^m g_{i,j}^m}{Z_{i,j}^m},
\label{eq:sinr_ntn}
\end{equation}
with the interference-plus-noise power
\begin{align}
Z_{i,j}^m ={}& \underbrace{\xi_{\NTN}\!\sum_{j'\neq j}E_{i,j'}^m P_{i,j'}^m g_{i,j}^m}_{\text{intra-beam}}
+\underbrace{\sum_{k\in\mathcal{K}}\sum_{l\in\mathcal{L}_k}V_{k,l}^m P_{k,l}^m g_{k,j}^m}_{\text{cross-tier TN-to-NTN}} \nonumber\\
&+\underbrace{\sum_{i'\neq i}\sum_{j'\in\mathcal{J}_{i'}}\!E_{i',j'}^m P_{i',j'}^m g_{i',j}^m+\sum_{b\in\mathcal{B}}\bar{P}^m g_{b,j}^m}_{\text{NTN inter-beam}}+\sigma^2.
\label{eq:interf_ntn}
\end{align}
The user throughputs are the Shannon rates
\begin{align}
R_{k,l} &= B_{\mathrm{RB}}\sum_{m\in\mathcal{M}}V_{k,l}^m\log_2\!\big(1+\sinr_{k,l}^{\TN}(m)\big), \label{eq:rate_tn}\\
R_{i,j} &= B_{\mathrm{RB}}\sum_{m\in\mathcal{M}}E_{i,j}^m\log_2\!\big(1+\sinr_{i,j}^{\NTN}(m)\big). \label{eq:rate_ntn}
\end{align}

\subsection{Resource-Allocation Problem Formulation}
\label{subsec:problem_formulation}

With $\mathcal{L}_k$ and $\mathcal{J}_i$ fixed by the offloading decision, the network chooses the powers $\mathbf{P}=\{P_{k,l}^m,P_{i,j}^m\}$ and the scheduling indicators $\boldsymbol{V}=\{V_{k,l}^m\}$ and $\boldsymbol{E}=\{E_{i,j}^m\}$. Sum-rate maximization, as in \cite{rahman2025joint}, tends to switch off users with poor channels. We instead maximize a utility $U(\cdot)$ of the user throughputs, which gives the problem
{\allowdisplaybreaks
\begin{align}
&\mathcal{P}_0:\ \max_{\mathbf{P},\,\boldsymbol{V},\,\boldsymbol{E}}\ \sum_{k\in\mathcal{K}}\sum_{l\in\mathcal{L}_k}U(R_{k,l})+\sum_{i\in\mathcal{I}}\sum_{j\in\mathcal{J}_i}U(R_{i,j}) \label{eq:p0_objective}\\
&\ \text{s.t.}\ \ \textstyle\sum_{m\in\mathcal{M}_{\TN}}\sum_{l\in\mathcal{L}_k}V_{k,l}^mP_{k,l}^m\le P_{k,\max},\ \forall k, \tag{C1a}\label{eq:c1a}\\
&\phantom{\ \text{s.t.}\ \ }\textstyle\sum_{m\in\mathcal{M}_{\NTN}}\sum_{j\in\mathcal{J}_i}E_{i,j}^mP_{i,j}^m\le P_{i,\max},\ \forall i, \tag{C1b}\label{eq:c1b}\\
&\phantom{\ \text{s.t.}\ \ }\textstyle1\le\sum_{l}V_{k,l}^m\le\mu_{\TN}\ \forall k,\,m\in\mathcal{M}_{\TN}, \tag{C2a}\label{eq:c2a}\\
&\phantom{\ \text{s.t.}\ \ }\textstyle\sum_{m}V_{k,l}^m\ge1\ \forall k,\,l\in\mathcal{L}_k, \tag{C2b}\label{eq:c2b}\\
&\phantom{\ \text{s.t.}\ \ }\textstyle1\le\sum_{j}E_{i,j}^m\le\mu_{\NTN}\ \forall i,\,m\in\mathcal{M}_{\NTN}, \tag{C2c}\label{eq:c2c}\\
&\phantom{\ \text{s.t.}\ \ }\textstyle\sum_{m}E_{i,j}^m\ge1\ \forall i,\,j\in\mathcal{J}_i, \tag{C2d}\label{eq:c2d}\\
&\phantom{\ \text{s.t.}\ \ }P_{k,l}^m\ge0,\ V_{k,l}^m\in\{0,1\},\ P_{k,l}^m=0\ \text{if}\ V_{k,l}^m=0, \tag{C3a}\label{eq:c3a}\\
&\phantom{\ \text{s.t.}\ \ }P_{i,j}^m\ge0,\ E_{i,j}^m\in\{0,1\},\ P_{i,j}^m=0\ \text{if}\ E_{i,j}^m=0. \tag{C3b}\label{eq:c3b}
\end{align}}
Constraints \eqref{eq:c1a}--\eqref{eq:c1b} enforce the power budgets of each \ac{gNB} and beam. Constraints \eqref{eq:c2a}--\eqref{eq:c2d} place at least one and at most $\mu_{\TN}$ or $\mu_{\NTN}$ users on every \ac{PRB} of an active carrier ($m\in\mathcal{M}_{\TN}$ or $m\in\mathcal{M}_{\NTN}$), and give every user at least one \ac{PRB}. The lower bounds apply only to a transmitter that serves at least one user. If a transmitter has more than $\mu M$ users, the constraints are infeasible. Our scheduler then raises its order to $\lceil n/M\rceil$ for $n$ users, and they share \acp{PRB} in the power domain. We adopt $U(r)=\ln r$ by default, the \ac{PF} problem \cite{ye2013userassociation}, whereas $U(r)=r$ gives the sum-rate problem~\cite{rahman2025joint}.

\section{Proposed SINR-Aware Adaptive Traffic Offloading}
\label{sec:proposed_framework}

\Ac{SAATO} replaces the \ac{RSRP} threshold and the cell-edge ring of~\cite{rahman2025joint} with a rule weighing interference, load and time. The network estimates each user's expected throughput on each tier every $T_{\mathrm{m}}$ from the filtered \acp{SINR} in \ac{UE} reports and the current loads. It filters the reports, counts the loads, compares expected throughputs and applies hysteresis.

\subsection{Filtered Link Quality and Tier Loads}
\label{subsec:state_variables}

\subsubsection{Layer-3 Filtered SINR}
Each \ac{UE} $u$ evaluates two candidate links, from its strongest \ac{gNB} $k(u)$ and from the beam $i(u)$ covering it. For a candidate transmitter $s$, with $s=k(u)$ for $\mathrm{ti}=\TN$ and $s=i(u)$ for $\mathrm{ti}=\NTN$, the \ac{NR} \ac{SS-SINR} measurement sets its reference power against all other co-channel transmitters $z\in\mathcal{K}\cup\mathcal{I}\cup\mathcal{B}$ at full load, $\hat{W}_{\mathrm{ti},u}^m=\big[(P_{s,\max}/M)\,g_{s,u}^m\big]\big/\big[\sum_{z\neq s}(P_{z,\max}/M)\,g_{z,u}^m+\sigma^2\big]$, where $P_{b,\max}=P_{i,\max}$. The per-\ac{PRB} values $\hat{W}_{\mathrm{ti},u}^m(t)$ pass through the mean-capacity (effective-\ac{SINR}) mapping
\begin{equation}
\hat{W}_{\mathrm{ti},u}(t) = \exp_2\!\Big(\frac{1}{M}\sum_{m\in\mathcal{M}_\mathrm{ti}}\log_2\big(1+\hat{W}_{\mathrm{ti},u}^m(t)\big)\Big)-1,
\label{eq:eff_sinr}
\end{equation}
where $\mathrm{ti}\in\{\TN,\NTN\}$. For $q<1$, terrestrial interference hits only part of the satellite carrier, and the mapping accounts for this by reflecting the whole carrier rather than its worst \acp{PRB}.

Rician fading and interference from nearby \acp{gNB} make the mapped values fluctuate. They are smoothed in the dB domain by the \ac{L3} filter of TS~38.331~\cite{3gpp38331},
\begin{equation}
\bar{W}_{\mathrm{ti},u}(t) = (1-Y)\,\bar{W}_{\mathrm{ti},u}(t-T_{\mathrm{m}}) + Y\,\hat{W}^{\mathrm{dB}}_{\mathrm{ti},u}(t),
\label{eq:l3_filter}
\end{equation}
where $\mathrm{ti}\in\{\TN,\NTN\}$, $\hat{W}^{\mathrm{dB}}_{\mathrm{ti},u}$ denotes $\hat{W}_{\mathrm{ti},u}$ in dB, and $Y$ is the filter weight per $T_{\mathrm{m}}$. TS~38.331 defines the per-sample weight as $2^{-\beta_{\mathrm{f}}/4}$ per layer-1 period for a configured coefficient $\beta_{\mathrm{f}}$, and $Y=0.25$ at $T_{\mathrm{m}}=1$~s approximately matches $\beta_{\mathrm{f}}\approx17$ at a 200-ms layer-1 period. The slow satellite-motion trend survives because the filter memory spans a few seconds and a pass lasts minutes. The filter starts from the first measurement, $\bar{W}_{\mathrm{ti},u}(0)=\hat{W}^{\mathrm{dB}}_{\mathrm{ti},u}(0)$. The filtered $\bar{W}_{\mathrm{ti},u}$ is in dB and $\bar{W}^{\mathrm{lin}}_{\mathrm{ti},u}$ is its linear value.

\subsubsection{Cell and Beam Loads}
Cell $k$ has $n_k(t)=|\mathcal{L}_k(t)|$ users across $\mu_{\TN}$ spatial layers per \ac{PRB}, which gives a normalized load per layer of $\eta_{\TN,k}(t) = n_k(t)/\mu_{\TN}$. Beam $i$ with $n_i(t)=|\mathcal{J}_i(t)|$ users has the load $C_{\NTN,i}(t) = n_i(t)/\mu_{\NTN}$. Offloading requires a usable satellite path, captured by the indicator $\mathds{1}_{\mathrm{LOS}}(u,t)$, which equals one if $\theta_u(t)\ge\theta_{\min}$ and the link is in \ac{LOS}, and zero otherwise.

\subsection{Load- and Capacity-Aware Threshold Adaptation}
\label{subsec:threshold_adaptation}

A strong link on a crowded tier can still deliver less than a weaker link on an empty one, which is why a fixed margin on the filtered \acp{SINR} falls short: it ignores terrestrial load and satellite occupancy. The rule therefore compares expected throughput. A transmitter with $n$ users and $\mu$ layers per \ac{PRB} splits its $\mu M$ \ac{PRB}-layers equally, which gives each user $\min\{M,\mu M/n\}$ \acp{PRB} on average, as a fluid approximation. With $u$ taken on the tier in question, the expected throughputs, normalized to $B_{\mathrm{RB}}$, are
\begin{align}
\hat{R}_{\TN,u}(t) &= \min\Big\{M,\frac{\mu_{\TN}M}{n_{k(u)}(t)}\Big\}\log_2\!\big(1+\bar{W}^{\mathrm{lin}}_{\TN,u}(t)\big), \label{eq:rhat_tn}\\
\hat{R}_{\NTN,u}(t) &= \min\Big\{M,\frac{\mu_{\NTN}M}{n_{i(u)}(t)}\Big\}\log_2\!\big(1+\bar{W}^{\mathrm{lin}}_{\NTN,u}(t)\big). \label{eq:rhat_ntn}
\end{align}
The decision metric is the throughput differential in dB,
\begin{equation}
G_u(t) = 10\log_{10}\big(\hat{R}_{\NTN,u}(t)/\hat{R}_{\TN,u}(t)\big).
\label{eq:gain_metric}
\end{equation}
Because $\min\{M,\mu M/n\}=M/\max\{1,n/\mu\}$, the metric splits into a link term and two load terms,
\begin{equation}
\begin{aligned}
G_u ={}& \underbrace{10\log_{10}\frac{\log_2(1+\bar{W}^{\mathrm{lin}}_{\NTN,u})}{\log_2(1+\bar{W}^{\mathrm{lin}}_{\TN,u})}}_{\Delta\bar{S}_u}
+\underbrace{10\log_{10}\max\{1,\eta_{\TN,k}\}}_{\Delta_{\mathrm{load}}(\eta_{\TN,k})} \\
&-\underbrace{10\log_{10}\max\{1,C_{\NTN,i}\}}_{\Delta_{\mathrm{cap}}(C_{\NTN,i})}.
\label{eq:gain_split}
\end{aligned}
\end{equation}

Testing $G_u$ against a base offset $\Theta_{\mathrm{base}}$ in dB amounts to checking the spectral-efficiency differential $\Delta\bar{S}_u$ against the adaptive threshold
\begin{equation}
\begin{split}
\Theta_{\mathrm{thresh}}(k,i,t) ={}& \Theta_{\mathrm{base}} - \Delta_{\mathrm{load}}\big(\eta_{\TN,k}(t)\big)\\
&+ \Delta_{\mathrm{cap}}\big(C_{\NTN,i}(t)\big).
\end{split}
\label{eq:adaptive_threshold}
\end{equation}
In a congested cell each user gets a smaller share of the layers, and the load term $\Delta_{\mathrm{load}}$ lowers the threshold. The capacity term $\Delta_{\mathrm{cap}}$ raises it as the beam fills. At light load both terms vanish, since $\max\{1,\cdot\}$ equals one when the normalized load does not exceed one. Both terms are logarithmic because they follow from the \ac{PF} objective $\Phi=\sum_{k,l}\ln R_{k,l}+\sum_{i,j}\ln R_{i,j}$ of $\mathcal{P}_0$, with no tuning involved.

\begin{proposition}
\label{prop:pf}
Hold all \acp{SINR} and the association of every user other than $u\in\mathcal{L}_k$ fixed, and assume that each transmitter shares its \ac{PRB}-layers equally among its users. Let $n_k$ count the users of cell $k$ including $u$, let $n_i$ count the users of beam $i$ including $u$ after the move, and let $n_k\ge\mu_{\TN}+1$, which ensures that the cell keeps at least $\mu_{\TN}$ users after $u$ leaves. The change $\Delta\Phi_u$ caused by moving $u$ from cell $k$ to beam $i$ then satisfies
\begin{equation}
\Delta\Phi_u = \frac{\ln 10}{10}\,G_u + \mathrm{err}_u,
\label{eq:prop_pf}
\end{equation}
where $|\mathrm{err}_u|<1/(2\min\{n_k-1,\,n_i-1\})$ if $n_i\ge\mu_{\NTN}+1$, and $\tfrac12\le\mathrm{err}_u<1$ if $u$ would be the only user of the beam ($n_i=1$; here $\mu_{\NTN}=1$ and the two cases cover every $n_i$).
\end{proposition}
\begin{proof}
For a fixed association, an equal split of resources maximizes the sum of the logarithms of the rates of a transmitter's users \cite{ye2013userassociation}. A transmitter with $n\ge\mu$ users of spectral efficiencies $c_v$ gives each of them the rate $B_{\mathrm{RB}}c_v\mu M/n$ and thereby contributes $\sum_v\ln c_v+n\ln(B_{\mathrm{RB}}\mu M)-\psi(n)$ to $\Phi$, with $\psi(n)=n\ln n$. An empty beam contributes nothing, which is the same expression with $n=0$ and $\psi(0)=0$. Moving $u$ changes only the terms of $u$, of cell $k$ (from $n_k$ to $n_k-1$ users) and of beam $i$ (from $n_i-1$ to $n_i$ users), and the hypotheses keep both transmitters in the range where this expression holds. With $c_{\mathrm{ti},u}=\log_2(1+\bar{W}^{\mathrm{lin}}_{\mathrm{ti},u})$ and the identity $\psi(n)-\psi(n-1)=\ln n+p(n-1)$, where $p(n)=n\ln(1+1/n)$ and $p(0)=0$,
\begin{align*}
\Delta\Phi_u &= \ln\frac{c_{\NTN,u}\,\mu_{\NTN}M/n_i}{c_{\TN,u}\,\mu_{\TN}M/n_k}-\big[p(n_i-1)-p(n_k-1)\big]\\
&= \ln\big(\hat{R}_{\NTN,u}/\hat{R}_{\TN,u}\big)+\mathrm{err}_u = \tfrac{\ln10}{10}\,G_u+\mathrm{err}_u,
\end{align*}
where \eqref{eq:rhat_tn}--\eqref{eq:rhat_ntn} were used with $n\ge\mu$. This gives $\mathrm{err}_u=p(n_k-1)-p(n_i-1)$. The inequality $\ln(1+z)\ge z-z^2/2$ gives $p(n)\in[1-1/(2n),1)$ for $n\ge1$, which yields the stated bound on $|\mathrm{err}_u|$ when $n_i\ge2$. If $n_i=1$, then $\mathrm{err}_u=p(n_k-1)\in[1/2,1)$.
\end{proof}

Comparing expected throughputs is thus justified: by Proposition~\ref{prop:pf}, a positive $G_u$ signals a move that raises the network \ac{PF} utility, up to an error $\mathrm{err}_u$ that shrinks with load. For the first user of an empty beam, $\mathrm{err}_u>0$ and the rule errs toward the terrestrial tier. At a typical load of 11 users per cell, this bias is about 4, 1 and 0.6~dB for the first, second and third satellite user. The interference a move creates or removes in the other tier enters only through later reports and the interference leakage terms of Joint-FP.

For comparison, a heuristic ablation variant replaces $\Delta_{\mathrm{load}}$ and $\Delta_{\mathrm{cap}}$ of \eqref{eq:adaptive_threshold} by $\Delta_{\mathrm{load}}=\zeta_1\eta_{\mathrm{h}}^2$ and $\Delta_{\mathrm{cap}}=\zeta_2(e^{C_{\mathrm{h}}}-1)$ and applies them to a plain \ac{SINR} difference. Here, $\eta_{\mathrm{h}}=\min\{n_k/20,1\}$ and $C_{\mathrm{h}}=\min\{n_i/(\mu_{\NTN}M),1\}$ are normalized cell and beam loads, and $\zeta_1$ and $\zeta_2$ are hand-tuned weights in dB.

\subsection{Hysteresis-Based Offloading Decision}
\label{subsec:algorithm_offloading}

Even a well-chosen threshold flips users that sit close to it when measurements fluctuate. Let $s_u(t)\in\{\TN,\NTN\}$ be the tier serving $u$ at time $t$. To prevent this, the rule associates $u$ with the satellite if and only if $\mathds{1}_{\mathrm{LOS}}(u,t)=1$ and
\begin{equation}
G_u(t) >
\begin{cases}
\Theta_{\mathrm{base}}+H, & s_u(t-T_{\mathrm{m}})=\TN,\\[0.3em]
\Theta_{\mathrm{base}}-H, & s_u(t-T_{\mathrm{m}})=\NTN,
\end{cases}
\label{eq:hysteresis}
\end{equation}
where $H$ is the hysteresis half-width in dB. By \eqref{eq:gain_split} and \eqref{eq:adaptive_threshold}, the same test reads $\Delta\bar{S}_u(t)>\Theta_{\mathrm{thresh}}(k,i,t)\pm H$. By Proposition~\ref{prop:pf}, a move to the satellite needs $\Delta\Phi_u-\mathrm{err}_u>\tfrac{\ln10}{10}(\Theta_{\mathrm{base}}+H)$. The dead zone of width $2H$ is thus a hysteresis on $G_u$ and only roughly a utility-gain threshold, since $\mathrm{err}_u$ is large for the first satellite user.

The loads in \eqref{eq:rhat_tn}--\eqref{eq:rhat_ntn} depend on the association being decided. To handle this coupling, Algorithm~\ref{alg:offloading} settles the loads and the association together over $T_{\mathrm{p}}$ relaxation passes from the last association. The relaxation is heuristic. Each pass relaxes the load estimate toward the loads $\tilde{\eta}^{(\mathrm{ps})}$ and $\tilde{C}^{(\mathrm{ps})}$ of its tentative association,
\begin{equation}
\eta^{(\mathrm{ps}+1)}=(1-\rho)\,\eta^{(\mathrm{ps})}+\rho\,\tilde{\eta}^{(\mathrm{ps})},\qquad \rho\in(0,1],
\label{eq:load_relaxation}
\end{equation}
where $\rho$ is the damping weight, $\eta$ stands for $\eta_{\TN,k}$ and the superscript $(\mathrm{ps})$ indexes the pass, and likewise for $C_{\NTN,i}$. The damping prevents a two-cycle in which, with $\rho=1$, a large batch of users leaves a crowded cell and returns once it looks attractive again. Joint-FP receives the resulting sets. The allocation does not change the next association decision, because the measured \ac{SS-SINR} assumes full-load interferers at fixed equal power.

\begin{algorithm}[t]
\caption{Proposed SAATO Procedure (Every $T_{\mathrm{m}}$)}
\label{alg:offloading}
\footnotesize\begin{algorithmic}[1]
\REQUIRE $\hat{W}^m_{\mathrm{ti},u}(t)$, $\bar{W}_{\mathrm{ti},u}(t-T_{\mathrm{m}})$, $s_u(t-T_{\mathrm{m}})$, $\forall u$; $\Theta_{\mathrm{base}}$, $H$, $Y$, $\rho$, $T_{\mathrm{p}}$
\ENSURE $\mathcal{L}_k$, $\forall k\in\mathcal{K}$; $\mathcal{J}_i$, $\forall i\in\mathcal{I}$
\STATE $\forall u$: $\mathds{1}_{\mathrm{LOS}}(u,t)\leftarrow$ indicator of Section~\ref{subsec:state_variables}; $\hat{W}_{\mathrm{ti},u}\leftarrow$ \eqref{eq:eff_sinr}; $\bar{W}_{\mathrm{ti},u}\leftarrow$ \eqref{eq:l3_filter}
\STATE $\eta_{\TN,k}$, $C_{\NTN,i}\leftarrow$ loads of the previous association
\FOR{$\mathrm{ps}=1$ \TO $T_{\mathrm{p}}$}
\STATE $\mathcal{L}_k\leftarrow\emptyset$, $\forall k$; $\mathcal{J}_i\leftarrow\emptyset$, $\forall i$
\FOR{$u=1$ \TO $N_{\mathrm{u}}$, with $k=k(u)$ and $i=i(u)$}
\STATE $G_u\leftarrow$ \eqref{eq:rhat_tn}--\eqref{eq:gain_metric}, with $u$ counted on the tier it would join
\STATE \textbf{if} $\mathds{1}_{\mathrm{LOS}}(u,t)=1$ and \eqref{eq:hysteresis} holds \textbf{then} $\mathcal{J}_i\leftarrow\mathcal{J}_i\cup\{u\}$ \textbf{else} $\mathcal{L}_k\leftarrow\mathcal{L}_k\cup\{u\}$
\ENDFOR
\STATE $\tilde{\eta}_{\TN,k}\leftarrow|\mathcal{L}_k|/\mu_{\TN}$, $\tilde{C}_{\NTN,i}\leftarrow|\mathcal{J}_i|/\mu_{\NTN}$; update $\eta_{\TN,k}$, $C_{\NTN,i}$ by \eqref{eq:load_relaxation}
\ENDFOR
\end{algorithmic}
\end{algorithm}

\section{Joint Cross-Tier Resource Allocation}
\label{sec:solution}
With the serving tiers fixed, the question is how well each user is served. We show why the joint problem is hard, reformulate it with fractional programming, derive a power update that charges each transmitter for the cross-tier interference it causes, add \ac{PRB} pruning and combine these steps into one convergent algorithm.

\subsection{Cross-Tier Coupling and Problem Complexity}
\label{subsec:problem_reformulation}

Under orthogonal spectrum \cite{rahman2025joint}, $\mathcal{P}_0$ with fixed $\mathcal{L}_k$ and $\mathcal{J}_i$ decouples into one terrestrial and one satellite problem with no shared variables, whereas for $q>0$ the coupling in \eqref{eq:interf_tn} and \eqref{eq:interf_ntn} removes this split. The per-tier scheme of \cite{rahman2025joint}, our benchmark, then sees the other tier only as uncoordinated interference.

\begin{theorem}
\label{thm:np_hardness}
Let the channel gains be arbitrary and let the number $K$ of terrestrial cells be part of the input. Then $\mathcal{P}_0$ with $U(r)=r$ is NP-hard, even for $M=1$, $q=0$, $\mathcal{I}=\emptyset$ and one scheduled user per terrestrial cell. We make no hardness claim for $U(r)=\ln r$.
\end{theorem}
\begin{proof}
Consider the restricted instance with $q=0$, no satellite beam ($\mathcal{I}=\emptyset$, which makes \eqref{eq:c2c}--\eqref{eq:c2d} void), one \ac{PRB} per carrier ($M=1$) and one user $l_k$ per cell. Constraints \eqref{eq:c2a}--\eqref{eq:c2b} then fix $V_{k,l_k}^1=1$ for every $k$, and $\mathcal{P}_0$ reduces to
\begin{equation}
\max_{0\le P_k\le P_{k,\max}}\ \sum_{k\in\mathcal{K}}\log_2\!\bigg(1+\frac{P_kg_{k,l_k}}{\sum_{k'\neq k}P_{k'}g_{k',l_k}+\sigma^2}\bigg),
\label{eq:np_hard_reduction}
\end{equation}
which is sum-rate maximization over a $K$-user Gaussian interference channel with interference treated as noise. This is the one-tone sum-rate spectrum-management problem of \cite{luo2008dynamic}, which is NP-hard. A common noise power $\sigma^2$ costs no generality because user-specific noise can be absorbed into the gains. Because this instance is a special case of $\mathcal{P}_0$, $\mathcal{P}_0$ is NP-hard.
\end{proof}
In \eqref{eq:np_hard_reduction}, $P_k=P_{k,l_k}^1$ and $g_{k',l_k}=g_{k',l_k}^1$. Unless P\,=\,NP, no polynomial-time algorithm finds a global optimum for every instance. We therefore develop Joint-FP, which alternates among auxiliary variables, powers and scheduling indicators.

\subsection{Fractional-Programming Reformulation}
\label{subsec:fp_derivation}
Joint-FP reduces both utilities to a weighted sum rate, with weights of one for $U(r)=r$. For \ac{PF}, each iteration resets the weights to a regularized gradient of $\Phi$ at the current point, namely $w_{k,l}=1/\max\{\tilde{R}_{k,l},r_{\min}\}$ and $w_{i,j}=1/\max\{\tilde{R}_{i,j},r_{\min}\}$. Here, $\tilde{R}=R/B_{\mathrm{RB}}$ is the rate normalized to the \ac{PRB} bandwidth, and the floor $r_{\min}$, in units of $\tilde{R}$, prevents division by zero. Algorithm~\ref{alg:joint_fp} evaluates $\Phi=\sum\ln\max\{\tilde{R},r_{\min}\}$ (the utility above up to an additive constant and the floor) at initialization, acceptance and the stopping test, since a user can have zero rate. The weights use $\log_2$ units while \eqref{eq:wsr} uses $\ln$, which only scales the objective by the common constant $\ln 2$. With fixed weights, the objective is
\begin{equation}
\begin{aligned}
F={}&\sum_{m\in\mathcal{M}}\sum_{k\in\mathcal{K}}\sum_{l\in\mathcal{L}_k}w_{k,l}V_{k,l}^m\ln\big(1+\sinr_{k,l}^{\TN}(m)\big) \\
&+\sum_{m\in\mathcal{M}}\sum_{i\in\mathcal{I}}\sum_{j\in\mathcal{J}_i}w_{i,j}E_{i,j}^m\ln\big(1+\sinr_{i,j}^{\NTN}(m)\big).
\label{eq:wsr}
\end{aligned}
\end{equation}

The Lagrangian dual transform \cite{shen2018fp2} introduces one auxiliary variable per user and \ac{PRB}, $\mathbf{x}=\{x_{k,l}^m,x_{i,j}^m\}\ge0$, removing each logarithm and turning \eqref{eq:wsr} into
\begin{equation}
\begin{aligned}
&f_1 ={}\!\sum_{m,k,l}w_{k,l}V_{k,l}^m\bigg[\ln(1+x_{k,l}^m)-x_{k,l}^m+\frac{(1+x_{k,l}^m)P_{k,l}^mg_{k,l}^m}{P_{k,l}^mg_{k,l}^m+Z_{k,l}^m}\bigg] \\
&+\sum_{m,i,j}w_{i,j}E_{i,j}^m\bigg[\ln(1+x_{i,j}^m)-x_{i,j}^m +\frac{(1+x_{i,j}^m)P_{i,j}^mg_{i,j}^m}{P_{i,j}^mg_{i,j}^m+Z_{i,j}^m}\bigg],
\label{eq:f1_uncoupled}
\end{aligned}
\end{equation}
where $Z_{k,l}^m$ and $Z_{i,j}^m$ come from \eqref{eq:interf_tn} and \eqref{eq:interf_ntn}. Setting $\partial f_1/\partial x=0$ gives the maximizers $x_{k,l}^{m\star}=\sinr_{k,l}^{\TN}(m)$ and $x_{i,j}^{m\star}=\sinr_{i,j}^{\NTN}(m)$, where $f_1=F$.

The problem remains non-convex in $\mathbf{P}$ because each bracket of \eqref{eq:f1_uncoupled} contains a ratio of powers. Applying the quadratic transform \cite{shen2018fp} with a second auxiliary variable $\mathbf{y}=\{y_{k,l}^m,y_{i,j}^m\}$ removes this ratio and gives
\begin{equation}
\begin{aligned}
&f_2 ={} \sum_{m\in\mathcal{M}}\sum_{k\in\mathcal{K}}\sum_{l\in\mathcal{L}_k}\Big[w_{k,l}V_{k,l}^m\big(\ln(1+x_{k,l}^m)-x_{k,l}^m\big)+2y_{k,l}^m\\
&\sqrt{w_{k,l}(1+x_{k,l}^m)\,V_{k,l}^mP_{k,l}^mg_{k,l}^m}-(y_{k,l}^m)^2\big(V_{k,l}^mP_{k,l}^mg_{k,l}^m+Z_{k,l}^m\big)\Big]\\
&+\sum_{m\in\mathcal{M}}\sum_{i\in\mathcal{I}}\sum_{j\in\mathcal{J}_i}\big[\,\cdots\,\big]_{(k,l,V)\to(i,j,E)},
\label{eq:f2_quadratic}
\end{aligned}
\end{equation}
where the last bracket mirrors the first for the satellite tier. The function $f_2$ is concave in $\mathbf{P}$ for fixed $\mathbf{x}$ and $\mathbf{y}$ and satisfies $f_1=\max_{\mathbf{y}}f_2$. The optimal $\mathbf{y}$ follows in closed form from $\partial f_2/\partial y=0$,
\begin{equation}
y_{k,l}^{m\star}=\frac{\sqrt{w_{k,l}(1+x_{k,l}^m)\,V_{k,l}^mP_{k,l}^mg_{k,l}^m}}{V_{k,l}^mP_{k,l}^mg_{k,l}^m+Z_{k,l}^m},
\label{eq:opt_y}
\end{equation}
and $y_{i,j}^{m\star}$ is analogous.

\subsection{Power Allocation With Cross-Tier Interference Leakage}
\label{subsec:power_update}
The quadratic form makes the power update tractable. After substituting \eqref{eq:interf_tn} and \eqref{eq:interf_ntn} into the penalties of \eqref{eq:f2_quadratic} and swapping the sums, each power appears in a single coefficient,
\begin{align}
&\sum_{m,k,l}(y_{k,l}^m)^2\big(V_{k,l}^mP_{k,l}^mg_{k,l}^m+Z_{k,l}^m\big)+\sum_{m,i,j}(y_{i,j}^m)^2\big(\cdots\big) \nonumber\\
&\quad=\sum_{m,k,l}V_{k,l}^mD_{k,l}^mP_{k,l}^m+\sum_{m,i,j}E_{i,j}^mD_{i,j}^mP_{i,j}^m+Q_0,
\label{eq:q_collected}
\end{align}
where $Q_0$ holds the background-beam and noise terms, free of $\mathbf{P}$. The coefficients are the interference leakage terms, and their last summands are the cross-tier leakage,
\begin{align}
D_{k,l}^m ={}& (y_{k,l}^m)^2g_{k,l}^m+\underbrace{\xi_{\TN}\sum_{l'\neq l}(y_{k,l'}^m)^2g_{k,l'}^m}_{\text{intra-cell}} \nonumber\\
&+\underbrace{\sum_{k'\neq k}\sum_{l'\in\mathcal{L}_{k'}}(y_{k',l'}^m)^2g_{k,l'}^m}_{\text{inter-cell}}
+\underbrace{\sum_{i\in\mathcal{I}}\sum_{j\in\mathcal{J}_i}(y_{i,j}^m)^2g_{k,j}^m}_{\text{cross-tier leakage}},
\label{eq:denom_tn}\\
D_{i,j}^m ={}& (y_{i,j}^m)^2g_{i,j}^m+\underbrace{\xi_{\NTN}\sum_{j'\neq j}(y_{i,j'}^m)^2g_{i,j'}^m}_{\text{intra-beam}} \nonumber\\
&+\underbrace{\sum_{i'\neq i}\sum_{j'\in\mathcal{J}_{i'}}(y_{i',j'}^m)^2g_{i,j'}^m}_{\text{inter-beam}}
+\underbrace{\sum_{k\in\mathcal{K}}\sum_{l\in\mathcal{L}_k}(y_{k,l}^m)^2g_{i,l}^m}_{\text{cross-tier leakage}},
\label{eq:denom_ntn}
\end{align}
where $g_{k,l'}^m$, $g_{i,j'}^m$, $g_{k,j}^m$ and $g_{i,l}^m$ are gains from the transmitter whose power is varied to the receivers it interferes with. The last terms of \eqref{eq:denom_tn} and \eqref{eq:denom_ntn} charge each transmitter, weighted by $(y_v^m)^2$, for the interference it causes in the other tier, and per-tier \ac{FP} lacks exactly these terms.

By \eqref{eq:q_collected}, $f_2$ splits per transmitter once $\mathbf{x}$, $\mathbf{y}$ and scheduling are fixed. Each transmitter $\mathrm{tx}\in\mathcal{K}\cup\mathcal{I}$ (the index sets are tagged, which keeps a gNB and a beam with the same number distinct) maximizes the concave function $\sum_{v,m}\big(2\sqrt{a_v^mP_v^m}-D_v^mP_v^m\big)$ under \eqref{eq:c1a} or \eqref{eq:c1b} with budget $P_{\mathrm{tx},\max}$ ($=P_{k,\max}$ or $P_{i,\max}$). Here $v$ runs over the users of $\mathrm{tx}$ and abbreviates $(k,l)$ or $(i,j)$ in every quantity, and $a_v^m=w_v(1+x_v^m)(y_v^m)^2g_v^m$. With a multiplier $\lambda_{\mathrm{tx}}\ge0$ for the budget, the \ac{KKT} conditions are necessary and sufficient, because the objective is concave and the constraint is linear, and they give the exact update that the proof of Theorem~\ref{thm:convergence} needs.

\begin{lemma}
\label{lem:power}
The optimal powers of transmitter $\mathrm{tx}$ are
\begin{equation}
P_v^{m\star}=\frac{a_v^m}{\big(D_v^m+\lambda_{\mathrm{tx}}\big)^2},
\label{eq:p_update}
\end{equation}
i.e., $P_{k,l}^{m\star}=a_{k,l}^m/(D_{k,l}^m+\lambda_k)^2$ for a \ac{gNB} and $P_{i,j}^{m\star}=a_{i,j}^m/(D_{i,j}^m+\lambda_i)^2$ for a beam. If $\sum_{v,m}a_v^m/(D_v^m)^2\le P_{\mathrm{tx},\max}$, the multiplier is $\lambda_{\mathrm{tx}}=0$. Otherwise, $\lambda_{\mathrm{tx}}$ is the unique root of $\sum_{v,m}a_v^m/(D_v^m+\lambda_{\mathrm{tx}})^2=P_{\mathrm{tx},\max}$ in $\big(0,\sqrt{\sum_{v,m}a_v^m/P_{\mathrm{tx},\max}}\,\big]$, found by bisection.
\end{lemma}
\begin{proof}
Stationarity of the Lagrangian $\sum_{v,m}(2\sqrt{a_v^mP_v^m}-D_v^mP_v^m)-\lambda_{\mathrm{tx}}\big(\sum_{v,m}P_v^m-P_{\mathrm{tx},\max}\big)$ gives $\sqrt{a_v^m/P_v^m}=D_v^m+\lambda_{\mathrm{tx}}$, which is \eqref{eq:p_update}. The budget sum is continuous and strictly decreasing in $\lambda_{\mathrm{tx}}$, and it is at most $P_{\mathrm{tx},\max}$ at the upper end of the interval. Users with $y_v^m=0$ have $a_v^m=0$ and $P_v^{m\star}=0$ and are skipped, and every remaining term has $D_v^m\ge(y_v^m)^2g_v^m>0$ by \eqref{eq:denom_tn} or \eqref{eq:denom_ntn}. The root therefore exists and is unique. Complementary slackness selects $\lambda_{\mathrm{tx}}=0$ when the unconstrained solution already meets the budget.
\end{proof}

\subsection{Interference-Aware PRB Scheduling}
\label{subsec:rbg_scheduling}
Power control alone cannot remove a user--\ac{PRB} pair that does more harm than good. The scheduler therefore prunes such pairs. It begins with a round-robin allocation satisfying \eqref{eq:c2a}--\eqref{eq:c2d}, where each \ac{gNB} co-schedules up to $\mu_{\TN}$ users per \ac{PRB} and each beam up to $\mu_{\NTN}$. Joint-FP then prunes the user--\ac{PRB} pairs that reduce the weighted sum rate. Switching off a terrestrial pair $(k,l,m)$ changes \eqref{eq:wsr} by
\begin{align}
\Delta_{k,l}^m ={}& \sum_{(k',l')\neq(k,l)}w_{k',l'}V_{k',l'}^m\ln\frac{1+\sinr_{k',l'}^{\TN,(-k,l)}(m)}{1+\sinr_{k',l'}^{\TN}(m)} \nonumber\\
&+\sum_{i\in\mathcal{I}}\sum_{j\in\mathcal{J}_i}w_{i,j}E_{i,j}^m\ln\frac{1+\sinr_{i,j}^{\NTN,(-k,l)}(m)}{1+\sinr_{i,j}^{\NTN}(m)} \nonumber\\
&-w_{k,l}\ln\big(1+\sinr_{k,l}^{\TN}(m)\big),
\label{eq:prune}
\end{align}
where $(-k,l)$ denotes an \ac{SINR} computed without $V_{k,l}^mP_{k,l}^m$ in the denominator, and $\Delta_{i,j}^m$ is defined likewise for a satellite pair. In Algorithm~\ref{alg:joint_fp}, $\Delta_v^m$ stands for $\Delta_{k,l}^m$ or $\Delta_{i,j}^m$. Removing a pair with $\Delta>0$ gains more at the other receivers of both tiers than it costs the pair itself. Each iteration removes at most one such pair per user while \eqref{eq:c2a}--\eqref{eq:c2d} hold. The batch of removals is accepted only if \eqref{eq:wsr} does not decrease. Otherwise only the single best pair is removed.

\subsection{Joint-FP Algorithm, Convergence and Complexity}
\label{subsec:algorithm_fp}

\begin{figure*}[!t]
\centering
\IfFileExists{figures/saato_jointfp_block.pdf}{%
\includegraphics[width=\textwidth]{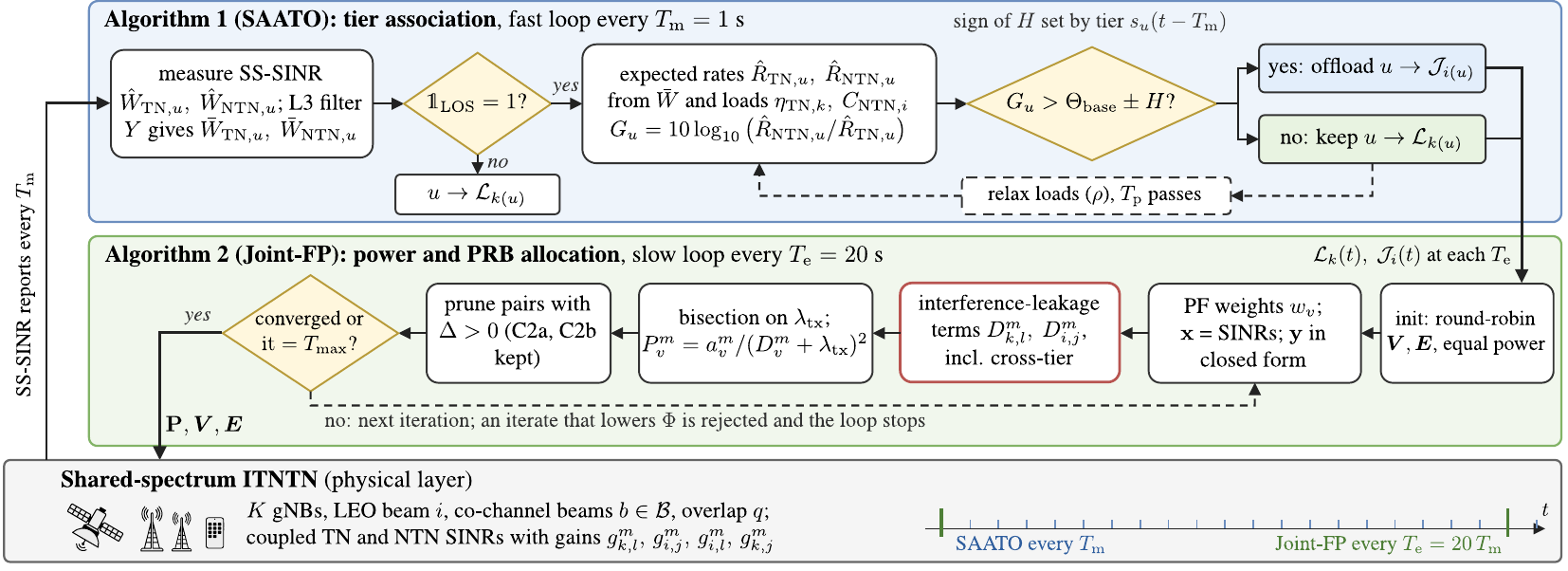}}{%
\fbox{\parbox[c][5.5cm][c]{0.9\textwidth}{\centering Export figures/saato\_jointfp\_block.drawio to PDF.}}}
\caption{Two-timescale operation: \acs{SAATO} (Algorithm~\ref{alg:offloading}) every $T_{\mathrm{m}}$ and Joint-FP (Algorithm~\ref{alg:joint_fp}) every $T_{\mathrm{e}}=20\,T_{\mathrm{m}}$.}
\label{fig:framework}
\end{figure*}

The auxiliary-variable updates of $\mathbf{x}$ and $\mathbf{y}$, the power update of Lemma~\ref{lem:power} and the pruning step of Section~\ref{subsec:rbg_scheduling} share one loop of Algorithm~\ref{alg:joint_fp}, which runs on the current association (Fig.~\ref{fig:framework}). Here $\mathrm{it}$ counts iterations and $\Phi$ is the utility being maximized, namely the floored logarithmic utility above for \ac{PF} and $F$ for the sum rate. The loop ends when $\Phi$ changes by less than $\mathrm{tol}$ in relative terms, when $\mathrm{it}$ reaches $T_{\max}$, or when an iteration would lower $\Phi$ (line~9).

\begin{algorithm}[!t]
\caption{Proposed Joint-FP Procedure (Every $T_{\mathrm{e}}$)}
\label{alg:joint_fp}
\footnotesize\begin{algorithmic}[1]
\REQUIRE $\mathcal{L}_k$, $\mathcal{J}_i$; all gains $g_{t,u}^m$, $t\in\mathcal{K}\cup\mathcal{I}\cup\mathcal{B}$; $P_{k,\max}$, $P_{i,\max}$; $\mathrm{tol}$, $T_{\max}$, $r_{\min}$
\ENSURE $\mathbf{P}$, $\boldsymbol{V}$, $\boldsymbol{E}$
\STATE $(\boldsymbol{V},\boldsymbol{E})\leftarrow$ round robin; $\mathbf{P}\leftarrow$ equal split of each budget; $\mathrm{it}\leftarrow0$; $\Phi^{(0)}\leftarrow\Phi(\mathbf{P},\boldsymbol{V},\boldsymbol{E})$
\REPEAT
\STATE $\mathrm{it}\leftarrow \mathrm{it}+1$; $w_v\leftarrow1/\max\{\tilde{R}_v,r_{\min}\}$ (\ac{PF}) or $w_v\leftarrow1$ (sum rate)
\STATE $x_v^m\leftarrow\sinr_v(m)$; $y_v^m\leftarrow$ \eqref{eq:opt_y}
\STATE $D_v^m\leftarrow$ \eqref{eq:denom_tn}--\eqref{eq:denom_ntn}; $a_v^m\leftarrow w_v(1+x_v^m)(y_v^m)^2g_v^m$
\STATE $\lambda_{\mathrm{tx}}\leftarrow$ bisection (Lemma~\ref{lem:power}); $P_v^m\leftarrow a_v^m/(D_v^m+\lambda_{\mathrm{tx}})^2$
\STATE $(\boldsymbol{V},\boldsymbol{E})\leftarrow$ remove pairs with $\Delta_v^m>0$ in \eqref{eq:prune}, s.t. \eqref{eq:c2a}--\eqref{eq:c2d}, if $F$ does not decrease, else only the best pair
\STATE $\Phi^{(\mathrm{it})}\leftarrow\Phi(\mathbf{P},\boldsymbol{V},\boldsymbol{E})$
\STATE \textbf{if} $\Phi^{(\mathrm{it})}<\Phi^{(\mathrm{it}-1)}$ \textbf{then} restore iterate $\mathrm{it}-1$ and \textbf{stop}
\UNTIL{$|\Phi^{(\mathrm{it})}-\Phi^{(\mathrm{it}-1)}|\le\mathrm{tol}|\Phi^{(\mathrm{it}-1)}|$ \OR $\mathrm{it}=T_{\max}$}
\end{algorithmic}
\end{algorithm}

\begin{theorem}
\label{thm:convergence}
With fixed weights (sum rate), the objective $F^{(\mathrm{it})}$ generated by Algorithm~\ref{alg:joint_fp} is non-decreasing. With \ac{PF} weights, the sequence of utilities $\Phi^{(\mathrm{it})}$ accepted by Algorithm~\ref{alg:joint_fp} is non-decreasing by the acceptance test. In both cases, the sequence is bounded above and converges.
\end{theorem}
\begin{proof}
Fix the weights of iteration $\mathrm{it}$. After the $\mathbf{x}$ and $\mathbf{y}$ updates, $f_2$ coincides with $F$ at the current point, because both updates are exact maximizers. With $\mathbf{P}^{(\mathrm{it}+\frac12)}$ the powers after Lemma~\ref{lem:power},
\begin{align*}
F\big(\mathbf{P}^{(\mathrm{it})},\boldsymbol{V}^{(\mathrm{it})},\boldsymbol{E}^{(\mathrm{it})}\big)
&= f_2\big(\mathbf{P}^{(\mathrm{it})},\boldsymbol{V}^{(\mathrm{it})},\boldsymbol{E}^{(\mathrm{it})},\mathbf{x}^\star,\mathbf{y}^\star\big)\\
&\le f_2\big(\mathbf{P}^{(\mathrm{it}+\frac12)},\boldsymbol{V}^{(\mathrm{it})},\boldsymbol{E}^{(\mathrm{it})},\mathbf{x}^\star,\mathbf{y}^\star\big)\\
&\le F\big(\mathbf{P}^{(\mathrm{it}+\frac12)},\boldsymbol{V}^{(\mathrm{it})},\boldsymbol{E}^{(\mathrm{it})}\big)\\
&\le F\big(\mathbf{P}^{(\mathrm{it}+1)},\boldsymbol{V}^{(\mathrm{it}+1)},\boldsymbol{E}^{(\mathrm{it}+1)}\big).
\end{align*}
The first inequality holds because Lemma~\ref{lem:power} maximizes $f_2$ exactly over the power block, the second because $F=\max_{\mathbf{x},\mathbf{y}}f_2$, and the third because a pruning step is accepted only if it does not decrease $F$. With \ac{PF} weights, this chain holds for the weights of iteration $\mathrm{it}$. The weights change between iterations, however, and the chain does not by itself order successive values of $\Phi$. The acceptance test in line~9 closes this gap: it accepts an iterate only if it does not lower $\Phi$. Every \ac{SINR} is bounded because the powers are bounded by the budgets and $\sigma^2>0$, and both sequences are bounded above as a result. Being monotone, they converge.
\end{proof}

Theorem~\ref{thm:convergence} concerns the utility sequence only; it establishes neither convergence of the iterates nor any form of optimality. With the schedule and weights fixed, the power iteration is the \ac{FP} iteration of \cite{shen2018fp, shen2018fp2}, and its fixed points meet the first-order conditions of the weighted sum rate. Forming each transmitter's aggregate power first lets one iteration evaluate all \acp{SINR} at cost $\mathcal{O}\big((K+I+|\mathcal{B}|)\,N_{\mathrm{u}}M_{\cup}\big)$. All interference leakages cost $\mathcal{O}\big((K+I)\,N_{\mathrm{u}}M_{\cup}\big)$, and the $N_{\mathrm{bis}}$ bisection iterations for $\lambda_{\mathrm{tx}}$ add $\mathcal{O}(N_{\mathrm{bis}}N_{\mathrm{u}}M_{\cup})$. The exact pruning test \eqref{eq:prune} dominates at $\mathcal{O}(N_{\mathrm{u}}^2M_{\cup})$. Algorithm~\ref{alg:offloading} costs $\mathcal{O}(T_{\mathrm{p}}N_{\mathrm{u}})$ per decision period $T_{\mathrm{m}}$, which is negligible next to Joint-FP. Joint-FP adds signaling: each transmitter needs its gains to receivers of the other tier and their $(y_v^m)^2$. The gains come from neighbor measurement reports exchanged once per $T_{\mathrm{e}}$, far longer than the 8~ms round trip. We assume the gains are available when the allocation is made, which makes our results optimistic for delayed or averaged gains.

\section{Simulation Results}
\label{sec:simulation}

\subsection{Simulation Setup and Benchmarks}
\label{subsec:setup}

\begin{table}[!t]
\centering
\caption{Simulation Parameters}
\label{tab:params}
\footnotesize
\renewcommand{\arraystretch}{1.0}
\setlength{\tabcolsep}{2.5pt}
\resizebox{\columnwidth}{!}{%
\begin{tabular}{|l|c|c|}
\hline
\textbf{Parameter} & \textbf{Symbol} & \textbf{Value} \\
\hline
Carrier frequency; bandwidth & $f_c$; $B$ & 2~GHz; 20~MHz \\ \hline
\acs{SCS}; \acsp{PRB} per carrier & $\Delta f$; $M$ & 30~kHz; 51 \\ \hline
Spectral overlap & $q$ & 0, 0.25, 0.5, 0.75, 1 \\ \hline
Noise density; noise figure & $N_0$; $N\!F$ & $-174$~dBm/Hz; 7~dB \\ \hline
\acs{PRB} noise; \acs{UE} antenna gain & $\sigma^2$; $G_{\mathrm{r}}$ & $-111$~dBm; 0~dBi \\ \hline
\acsp{gNB}; cell radius & $K$; $R_{\mathrm{cell}}$ & 7; 1~km \\ \hline
Users/cell; fringe; annulus & -- & 6--16; 30; 2.7--6~km \\ \hline
\acs{gNB} budget; antenna gain & $P_{k,\max}$; $G_{\mathrm{t}}$ & 46~dBm; 20~dBi \\ \hline
\acs{TN} Rician factor; coh.\ block & $K_{\mathrm{R},\TN}$; $N_{\mathrm{coh}}$ & 9~dB; 12 \acsp{PRB} \\ \hline
\acs{MU-MIMO} order; leakage & $\mu_{\TN}$; $\xi_{\TN}$ & 5; $-15$~dB \\ \hline
Altitude; \acs{EIRP} density & -- & 600~km; 34~dBW/MHz \\ \hline
Satellite gain; beamwidth & $G_{\mathrm{s}}$; $\mathrm{ang}_{3\mathrm{dB}}$ & 30~dBi; $4.41^\circ$ \\ \hline
Beam budget; center signal & $P_{i,\max}$; -- & 47~dBm; $-100$~dBm \\ \hline
Pol., scint., atm.\ loss & $L_{\mathrm{pol}}$; $L_{\mathrm{sc}}$; $L_{\mathrm{atm}}$ & 3; 2.2; 0.1~dB \\ \hline
Satellite Rician factor & $K_{\mathrm{R},\NTN}$ & 10~dB \\ \hline
\acs{NTN} users per \acs{PRB}; leakage & $\mu_{\NTN}$; $\xi_{\NTN}$ & 1; 0~dB \\ \hline
Serving; co-channel beams & $I$; $|\mathcal{B}|$ & 1; 6 \\ \hline
Peak-elev.\ bound; min.\ elev. & $\theta_{\mathrm{pk}}$; $\theta_{\min}$ & $50^\circ$; $30^\circ$ \\ \hline
Decision; allocation period & $T_{\mathrm{m}}$; $T_{\mathrm{e}}$ & 1~s; 20~s \\ \hline
\acs{L3} filter weight per $T_{\mathrm{m}}$ & $Y$ & 0.25 \\ \hline
Base offset; hysteresis & $\Theta_{\mathrm{base}}$; $H$ & 0~dB; 1~dB \\ \hline
Relaxation passes; damping & $T_{\mathrm{p}}$; $\rho$ & 5; 0.4 \\ \hline
\acs{RSRP} threshold; edge ring & $\mathrm{RSRP}_{\mathrm{th}}$; -- & $-100$~dBm/RE; 200~m \\ \hline
Fixed threshold; heuristic weights & $\mathrm{Th}_{\mathrm{fix}}$; $\zeta_1$, $\zeta_2$ & 0~dB; 6~dB each \\ \hline
Joint-FP iterations; tolerance & $T_{\max}$; $\mathrm{tol}$ & 30; $10^{-4}$ \\ \hline
Joint-FP rate floor & $r_{\min}$ & $10^{-3}$~b/s/Hz \\ \hline
Ping-pong window; outage limit & -- & 5~s; 0.5~Mbps \\ \hline
Drops; snapshots per drop & -- & 30; 11--12 \\
\hline
\end{tabular}}
\end{table}

The terrestrial setup follows \cite{rahman2025joint} (parameters in Table~\ref{tab:params}), except for the carrier numerology of Section~\ref{subsec:architecture}. The satellite tier uses the \ac{LEO}-600 S-band handheld scenario in TR~38.821 \cite{3gpp38821}. All schemes are evaluated on the same drops. Each drop uses a new user population and one satellite pass, with fading redrawn every $T_{\mathrm{m}}$. A drop contains 11 or 12 snapshots, one per $T_{\mathrm{e}}$, and each user--snapshot pair contributes one sample. At every snapshot, resource allocation uses the instantaneous per-\ac{PRB} channel conditions. Joint-FP thus uses genie-aided channel knowledge, and the hold over $T_{\mathrm{e}}$ is not simulated. Confidence intervals for network throughput use Student's $t$-intervals across drops, while the 5th percentile and the geometric mean are bootstrapped over whole drops.

We evaluate seven schemes to separate the effects of association, cross-tier power control and offloading. The first is the reference design of \cite{rahman2025joint}, denoted RSRP + per-tier FP, which offloads a user with a usable satellite link when its \ac{RSRP} per resource element (RE) falls below $\mathrm{RSRP}_{\mathrm{th}}$ or when the user lies in the edge ring of its cell. The second rule applies a fixed-\ac{SINR} bias, $\mathrm{Th}_{\mathrm{fix}}$, and offloads a user when the unfiltered \ac{SINR} difference exceeds it, without hysteresis or load awareness. The third is the proposed \ac{SAATO}. Each rule is paired with per-tier \ac{FP}, and Joint-FP is evaluated with the \ac{RSRP} rule and with \ac{SAATO}. Both allocators share the same \ac{PF} weights, initialization, pruning and iteration limit and differ only in how they treat cross-tier interference leakage. The last two schemes disable offloading, one with equal power and one with per-tier \ac{FP}. A tier switch occurs when a user changes its serving tier, a ping-pong when it switches back within the ping-pong window, and outage means user throughput below 0.5~Mbps.

\begin{figure*}[!t]
\centering
\includegraphics[width=172mm]{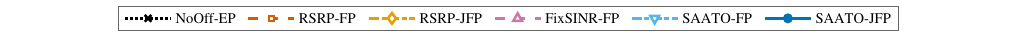}\\[-2pt]
\subfloat[]{\includegraphics[width=51mm]{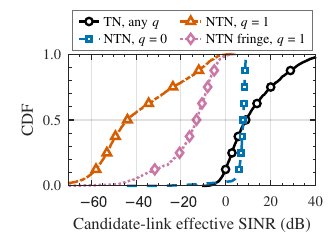}}\hfil
\subfloat[]{\includegraphics[width=51mm]{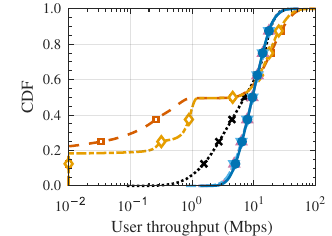}}\hfil
\subfloat[]{\includegraphics[width=51mm]{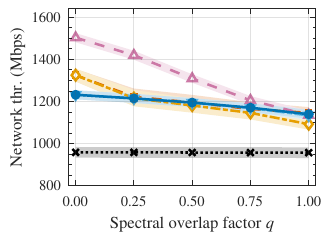}}\\[-2pt]
\subfloat[]{\includegraphics[width=51mm]{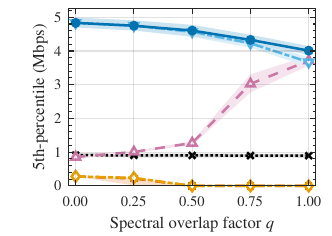}}\hfil
\subfloat[]{\includegraphics[width=51mm]{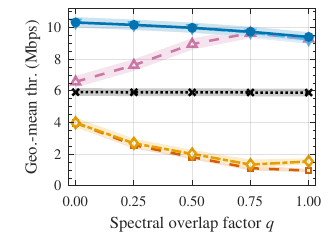}}\hfil
\subfloat[]{\includegraphics[width=51mm]{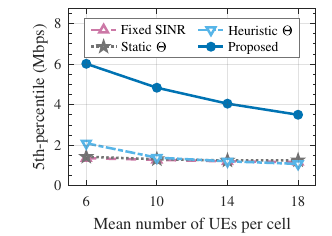}}
\caption{Results over 30 drops. (a) \ac{CDF} of the candidate-link \acs{SINR} at the pass mid-point. (b) \acs{CDF} of user throughput at $q=1$. (c) Network, (d) 5th-percentile and (e) geometric-mean throughput versus $q$, with 95\% confidence intervals. (f) 5th-percentile throughput versus terrestrial load at $q=0.5$. The legend applies to (b)--(e), while (a) and (f) have their own; NoOff-EP is no offloading with equal power, and -FP and -JFP denote per-tier FP and Joint-FP.}
\label{fig:res1}
\end{figure*}

\subsection{Cross-Tier SINR and Offloading Behavior}
\label{subsec:res_physics}

The satellite link provides competitive \ac{SINR} for a substantial share of users when overlap is limited, but this advantage diminishes under full overlap (Fig.~\ref{fig:res1}(a)). With adjacent carriers ($q=0$), the satellite \ac{SINR} for 90\% of users sits between about 4 and 9~dB, since a 50~km beam is almost uniform over 12~km. The terrestrial \ac{SINR} reaches 36~dB near the sites and falls to $-3$~dB at the cell edges, which makes the satellite at least as good for 44\% of users. Under full overlap, terrestrial interference dominates the satellite link and pulls the median satellite \ac{SINR} to about $-44$~dB, and even distant fringe users see $-13$~dB. Offloading therefore pays off only for some users and overlap levels: in this geometry, the satellite mainly serves the rural fringe and users on non-overlapping spectrum. 

Fig.~\ref{fig:res1}(b) compares the throughput CDFs at $q=1$ (samples below 0.01~Mbps are plotted at 0.01~Mbps). The \ac{RSRP} rule retains users with high throughput in the cells, but some of the users it offloads experience near-zero throughput: 22\% of all user-snapshots with per-tier \ac{FP} and 18\% with Joint-FP. \ac{SAATO} leaves almost every user on its better tier and gives a steep \acs{CDF}, with no outage sample among its 37\,333 user-snapshots from the 30 drops. The fixed-\ac{SINR} rule reaches a similar distribution here, largely because few users meet its offloading condition, but with many more tier switches (Section~\ref{subsec:res_mobility}). 

\begin{table*}[!t]
\centering
\caption{Performance Comparison at $q=0.5$ and $q=1$ (Bold Marks the Proposed Method, Not the Best Value)}
\label{tab:main}
\footnotesize
\renewcommand{\arraystretch}{1.15}
\setlength{\tabcolsep}{4.5pt}
\begin{tabular}{|l||c|c|c|c|c||c|c|c|c|c|}
\hline
\multirow{3}{*}{\textbf{Scheme}} & \multicolumn{5}{c||}{$\boldsymbol{q=0.5}$} & \multicolumn{5}{c|}{$\boldsymbol{q=1}$} \\
\cline{2-11}
 & Sum & 5\%-ile & Geo.\ mean & Outage & Switch rate & Sum & 5\%-ile & Geo.\ mean & Outage & Switch rate \\
 & (Mbps) & (Mbps) & (Mbps) & (\%) & (/user/min) & (Mbps) & (Mbps) & (Mbps) & (\%) & (/user/min) \\
\hline\hline
No offload, equal power & 959 & 0.91 & 5.91 & 1.3 & 0.000 & 958 & 0.90 & 5.90 & 1.4 & 0.000 \\ \hline
No offload, per-tier FP & 1126 & 3.91 & 9.28 & 0.0 & 0.000 & 1124 & 3.84 & 9.25 & 0.0 & 0.000 \\ \hline
RSRP + per-tier FP \cite{rahman2025joint} & 1196 & 0.00 & 1.82 & 37.8 & 0.009 & 1147 & 0.00 & 0.97 & 44.1 & 0.009 \\ \hline
RSRP + Joint-FP & 1181 & 0.00 & 2.02 & 34.4 & 0.009 & 1093 & 0.00 & 1.55 & 26.5 & 0.009 \\ \hline
Fixed-SINR + per-tier FP & 1309 & 1.27 & 8.96 & 0.1 & 6.252 & 1136 & 3.71 & 9.29 & 0.0 & 0.800 \\ \hline
SAATO + per-tier FP & 1195 & 4.57 & 9.98 & 0.0 & 0.040 & 1134 & 3.67 & 9.25 & 0.0 & 0.020 \\ \hline
\textbf{SAATO + Joint-FP (proposed)} & 1195 & \textbf{4.60} & \textbf{9.99} & \textbf{0.0} & 0.040 & 1140 & \textbf{4.01} & \textbf{9.41} & \textbf{0.0} & 0.020 \\
\hline
\end{tabular}
\end{table*}

Table~\ref{tab:main} compares all schemes at $q=0.5$ and $q=1$. The 95\% confidence intervals stay within $\pm36$~Mbps for network throughput and $\pm0.4$~Mbps for the other metrics. Although \ac{SAATO} does not use $q$ as an input, it offloads 8.1 users on average at $q=0$ but only 1.5 at $q=1$, because link conditions change with spectrum overlap. The \ac{RSRP} rule, by contrast, relies on geometry and offloads the same number of users (52.4 on average) at every overlap level. At full overlap, it sends those users to satellite links with \ac{SINR} well below 0~dB. The remaining terrestrial users benefit from the freed resources; the reference design therefore keeps a high network throughput despite its 44\% outage. What matters is therefore which users are offloaded, not how many. 

\subsection{Comparison With a Retuned RSRP Baseline}
\label{subsec:res_rsrp}

\begin{figure}[!t]
\centering
\IfFileExists{figures/plot_rsrp_fig.pdf}{%
\includegraphics[width=0.92\columnwidth]{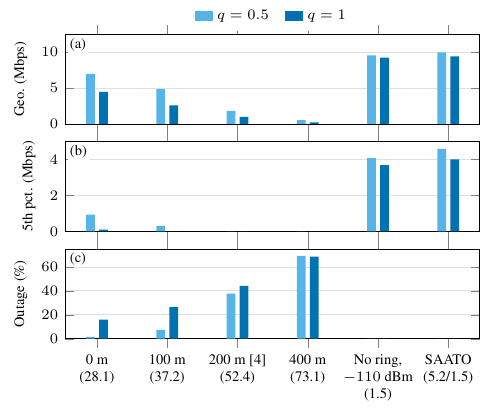}}{%
\fbox{\parbox[c][4cm][c]{0.9\columnwidth}{\centering Upload figures/plot\_rsrp\_fig.pdf}}}
\caption{Retuned \acs{RSRP} baseline over 30 drops: (a) geometric-mean throughput, (b) 5th-percentile throughput and (c) outage. Groups give the edge-ring width at $\mathrm{RSRP}_{\mathrm{th}}=-100$~dBm per RE, or no ring at $-110$~dBm; the last group is \acs{SAATO} with Joint-FP. Numbers in parentheses are the average offloaded users ($q=0.5$/$q=1$ for \acs{SAATO}). Missing bars are zero. The \acs{RSRP} bars use per-tier \acs{FP}, and the $-90$~dBm result is given in the text.}
\label{fig:rsrp}
\end{figure}

To test whether the \ac{RSRP} rule is mistuned for shared spectrum, we sweep $\mathrm{RSRP}_{\mathrm{th}}$ from $-110$ to $-90$~dBm per resource element in Fig.~\ref{fig:rsrp}. The outage rate changes by at most 1.8 percentage points, because most offloaded users are selected by the edge-ring condition or lie in the fringe. Changing the ring width has a larger effect, but even a ring of 0~m does not resolve the problem: at $q=1$, the rule still offloads 28.1 of the 30 fringe users, leaving 15.9\% of users in outage and a geometric-mean throughput of 4.48~Mbps, against 5.90~Mbps for no offloading with equal power. With no ring and $\mathrm{RSRP}_{\mathrm{th}}=-110$~dBm, it offloads just 1.5 users. With Joint-FP it then reaches 96\% and 100\% of the geometric-mean throughput of \ac{SAATO} with Joint-FP at $q=0.5$ and $q=1$ (96\% and 98\% with per-tier \ac{FP}, the setting plotted in Fig.~\ref{fig:rsrp}). This setting is chosen in hindsight: at $-90$~dBm, the same rule achieves only 7.28~Mbps at $q=1$. \ac{SAATO} needs no such tuning, because the weakness of the \ac{RSRP} rule lies in its information, not its threshold. 

\subsection{Effect of Spectral Overlap and the Role of Joint-FP}
\label{subsec:res_alpha}

We next vary the overlap from adjacent to fully shared carriers (Figs.~\ref{fig:res1}(c)--(e)). The outage rate of the \ac{RSRP} rule rises from 9.9\% at $q=0$ to 37.8\% at $q=0.5$ and 44.1\% at $q=1$, while its geometric-mean throughput stays below that of both no-offloading references. At full overlap, \ac{SAATO} with Joint-FP achieves a geometric-mean throughput about 60\% above the equal-power no-offloading reference, 2\% above no offloading with per-tier \ac{FP} and 9.7 times that of the \ac{RSRP}-based reference design. The fixed-\ac{SINR} rule behaves differently at low overlap. At $q=0$, it offloads about 41 users, compared with 8 for \ac{SAATO}, and gains 22\% in network throughput over \ac{SAATO}. This does not translate into better user throughput: its 5th-percentile throughput is 0.86~Mbps against 4.83~Mbps for \ac{SAATO}, and its geometric-mean throughput is 36\% lower. Its network-throughput advantage shrinks with overlap and is gone at $q=1$. 

With the \ac{RSRP} rule at $q=1$, Joint-FP reduces outage from 44.1\% to 26.5\%, a 40\% relative reduction, and raises geometric-mean throughput by 60\%, at the cost of 4.7\% of network throughput. The gain of Joint-FP is smaller under \ac{SAATO}, which uses link conditions rather than \ac{RSRP} alone to select users. At $q=0.5$, Joint-FP adds little, but at $q=1$ it increases the geometric-mean and 5th-percentile throughput by 0.16 and 0.34~Mbps over \ac{SAATO} with per-tier \ac{FP}, on common drops.

The no-offloading rows of Table~\ref{tab:main} help separate the gains from power control and offloading. At $q=0.5$, power control alone raises geometric-mean throughput from 5.91 to 9.28~Mbps. On top of that, \ac{SAATO} with Joint-FP adds 0.71~Mbps in geometric-mean throughput and 0.70~Mbps at the 5th percentile, increases of 8\% and 18\% with 95\% intervals of $[0.68, 0.75]$ and $[0.63, 0.76]$. At $q=1$, the additional gains are 0.16 and 0.17~Mbps, or 2\% and 4\%, with intervals of $[0.14, 0.19]$ and $[0.12, 0.21]$. Power control thus accounts for most of the gain, while offloading adds a smaller benefit when users are chosen by link conditions. 

\subsection{Switching Stability and Parameter Sensitivity}
\label{subsec:res_mobility}

\begin{figure*}[!t]
\centering
\subfloat[]{\includegraphics[width=84mm, height= 42mm]{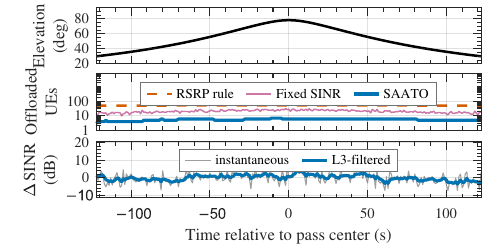}}\hfil
\subfloat[]{\includegraphics[width=84mm, height= 42mm]{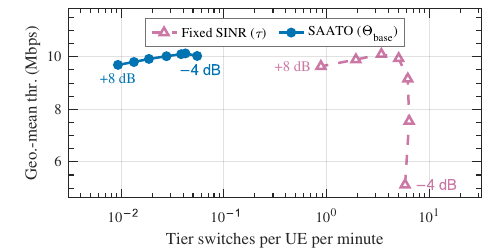}}
\caption{Switching behavior at $q=0.5$. (a) One \acs{LEO} pass: elevation, offloaded users and instantaneous and \acs{L3}-filtered \acs{SINR} differences of one user. (b) Geometric-mean throughput versus tier-switching rate for offsets $\mathrm{Th}_{\mathrm{fix}}$ or $\Theta_{\mathrm{base}}$ from $-4$ to $+8$~dB.}
\label{fig:res2}
\end{figure*}

\begin{figure*}[!t]
\centering
\subfloat[]{\includegraphics[width=51mm]{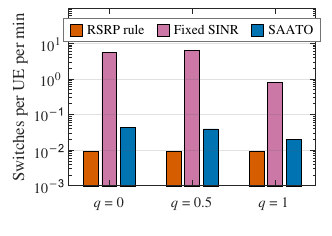}}\hfil
\subfloat[]{\includegraphics[width=51mm]{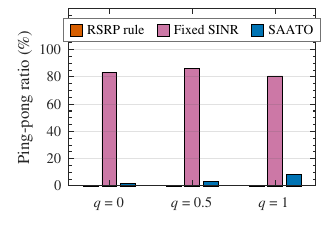}}\hfil
\subfloat[]{\includegraphics[width=51mm]{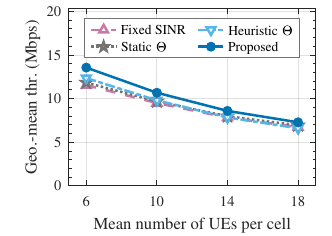}}\\[-2pt]
\subfloat[]{\includegraphics[width=51mm]{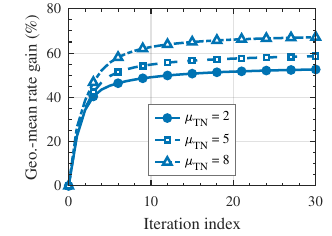}}\hfil
\subfloat[]{\includegraphics[width=51mm]{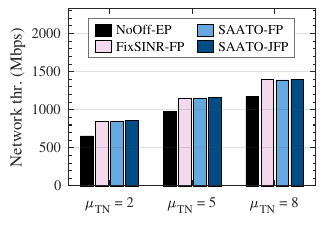}}\hfil
\subfloat[]{\includegraphics[width=51mm]{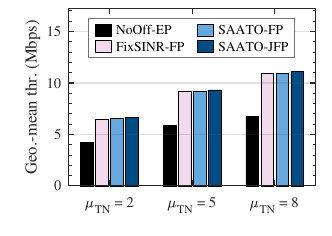}}
\caption{Stability, load, convergence and \acs{MU-MIMO} results. (a) Tier-switching rate and (b) ping-pong ratio. (c) Geometric-mean throughput versus terrestrial load at $q=0.5$. (d) Geometric-mean rate gain of Joint-FP over the equal-power start at $q=1$. (e) Network and (f) geometric-mean throughput versus the \acs{MU-MIMO} order at $q=1$ (separate sweep). Abbreviations as in Fig.~\ref{fig:res1}; Proposed is \acs{SAATO}.}
\label{fig:res3}
\end{figure*}

Because the fixed-\ac{SINR} rule responds to short-term channel fluctuations, its offloaded count jumps between 10 and 32 users during the pass in Fig.~\ref{fig:res2}(a), where the satellite peaks at about $78^\circ$ elevation. The instantaneous \ac{SINR} difference fluctuates by several decibels around its trend, whereas filtering and hysteresis keep the \ac{SAATO} count between three and seven. The same contrast appears across all drops in Figs.~\ref{fig:res3}(a) and (b). For $q\le0.5$, the fixed-\ac{SINR} rule makes 5.6--6.3 tier switches per user per minute, falling to 0.8 at $q=1$, and 81--86\% of them are ping-pongs. \ac{SAATO} makes 0.02--0.04 switches per user per minute, with at most 9\% ping-pongs. 

Fig.~\ref{fig:res2}(b) tests whether threshold selection alone can remove the drawbacks of the fixed-\ac{SINR} rule. At $-4$~dB, it offloads too many users and the geometric-mean throughput falls to 5.1~Mbps. Higher thresholds raise it to 9.6--10.1~Mbps, but even the best point needs about ninety times as many tier switches as \ac{SAATO}. By comparison, \ac{SAATO} maintains 9.7--10.1~Mbps and 0.009--0.055 switches per user per minute across all tested offsets and is therefore insensitive to $\Theta_{\mathrm{base}}$ at $q=0.5$. 

\begin{table}[!t]
\centering
\caption{Sensitivity of SAATO + Joint-FP at $q=0.5$}
\label{tab:sens}
\footnotesize
\renewcommand{\arraystretch}{1.05}
\setlength{\tabcolsep}{4pt}
\begin{tabular}{|c|c||c|c|c|c|c|}
\hline
\textbf{Param.} & \textbf{Value} & \textbf{Geo} & \textbf{5\%-ile} & \textbf{Out.} & \textbf{Switch} & \textbf{PP} \\
\hline\hline
\multirow{3}{*}{$H$ (dB)} & 0 & 9.99 & 4.58 & 0.0 & 0.371 & 45.3 \\
 & \textbf{1} & 9.99 & 4.60 & 0.0 & 0.040 & 3.1 \\
 & 3 & 9.93 & 4.51 & 0.0 & 0.007 & 0.0 \\ \hline
\multirow{3}{*}{$Y$} & 1 & 10.02 & 4.62 & 0.0 & 1.306 & 65.2 \\
 & \textbf{0.25} & 9.99 & 4.60 & 0.0 & 0.040 & 3.1 \\
 & 0.0625 & 9.97 & 4.57 & 0.0 & 0.011 & 0.7 \\ \hline
\multirow{3}{*}{$\rho$} & \textbf{0.4} & 9.99 & 4.60 & 0.0 & 0.040 & 3.1 \\
 & 0.7 & 9.98 & 4.60 & 0.0 & 0.055 & 6.5 \\
 & 1 & 3.06 & 0.00 & 25.9 & 27.05 & 99.1 \\ \hline
\multirow{4}{*}{$T_{\mathrm{p}}$} & 1 & 3.06 & 0.00 & 25.9 & 27.05 & 99.1 \\
 & 3 & 9.62 & 4.21 & 0.0 & 0.018 & 0.9 \\
 & \textbf{5} & 9.99 & 4.60 & 0.0 & 0.040 & 3.1 \\
 & 10 & 9.99 & 4.59 & 0.0 & 0.042 & 2.4 \\
\hline
\end{tabular}
\\[3pt]
\parbox{\columnwidth}{\scriptsize Defaults in bold. Geo is the geometric mean and 5\%-ile the 5th percentile, both in Mbps; outage (Out.) and ping-pong ratio (PP) are in \%, switch rate in tier switches per user per minute. $Y=1$ means no filtering.}
\end{table}
We next examine the other \ac{SAATO} parameters in Table~\ref{tab:sens}. Across the tested values of $H$ and $Y$, geometric-mean throughput stays within 9.93--10.02~Mbps; these parameters therefore matter for stability rather than throughput. This becomes clear when the stabilizing mechanisms are removed: dropping hysteresis multiplies the switching rate by nine, and removing the filter raises it to 1.3 switches per user per minute, about 33 times the default. The load-relaxation damping in \eqref{eq:load_relaxation} is also important. With $\rho=1$ or a single relaxation pass, the load estimate falls into the two-cycle of Section~\ref{subsec:algorithm_offloading}, which drives outage to 25.9\% and the switching rate to 27 switches per user per minute. For $\rho\in[0.4,0.7]$ and $T_{\mathrm{p}}\ge5$, throughput and outage stay at their defaults. 

\subsection{Terrestrial Load, Convergence and MU-MIMO Order}
\label{subsec:res_density}

Fig.~\ref{fig:res1}(f) examines the decision metric as the number of users per cell grows from 4--8 to 16--20. It compares the proposed metric with the heuristic threshold, a static threshold ($\zeta_1=\zeta_2=0$) with filtering and hysteresis only, and the fixed rule. The heuristic and static rules reach only 1.1--2.1~Mbps at the 5th percentile, against 3.5--6.0~Mbps for the proposed metric, which also achieves the highest geometric-mean throughput at every load (Fig.~\ref{fig:res3}(c)).
The proposed metric limits offloading as the satellite load grows, whereas the heuristic and static rules keep offloading until the beam becomes the bottleneck: they offload 13--27 users, against only 3.8--6.8 under the proposed metric. Under the proposed metric, each additional satellite user raises $\Delta_{\mathrm{cap}}$ for the next, a capacity loss that the logarithmic terms of Proposition~\ref{prop:pf} account for. At every load, the heuristic and static rules reach about 6--10\% higher network throughput by offloading more users, at the cost of lower 5th-percentile throughput. 

Fig.~\ref{fig:res3}(d) shows the Joint-FP utility over the iterations for 30 snapshots at $q=1$, one per drop. It never decreases, as Theorem~\ref{thm:convergence} guarantees; the panel therefore only shows how quickly it rises: the geometric-mean rate gain reaches 45--56\% above the equal-power start within five iterations. Network throughput of every scheme grows with the \ac{MU-MIMO} order, but the fairness ranking does not change (Figs.~\ref{fig:res3}(e) and (f)): \ac{SAATO} with Joint-FP keeps the highest or tied-highest geometric-mean throughput at every order. At $\mu_{\TN}=5$ and $q=1$, the reference design leaves 44.1\% of its samples in outage (Table~\ref{tab:main}), whereas \ac{SAATO} with Joint-FP leaves none. 

\subsection{Robustness of the Gains}
\label{subsec:res_robust}

\begin{table}[!t]
\centering
\caption{Robustness of the SAATO Gains (Paired Differences in Mbps)}
\label{tab:robust}
\scriptsize
\renewcommand{\arraystretch}{1.1}
\setlength{\tabcolsep}{2.2pt}
\begin{tabular}{|l||c|c||c|c||c|c|}
\hline
 & \multicolumn{2}{c||}{$q=0.5$} & \multicolumn{4}{c|}{$q=1$} \\
\cline{2-7}
 & \multicolumn{2}{c||}{Offloading} & \multicolumn{2}{c||}{Joint-FP} & \multicolumn{2}{c|}{Offloading} \\
\cline{2-7}
\textbf{Case} & Geo & 5\%-ile & Geo & 5\%-ile & Geo & 5\%-ile \\
\hline\hline
Fresh drops & +0.72 & +0.70 & +0.14 & +0.34 & +0.17 & +0.16 \\ \hline
Large-scale \acs{CSI} & +0.75 & +0.63 & +0.19 & +0.39 & +0.16 & +0.10 \\ \hline
4~dB \acs{CSI} error & +0.72 & +0.63 & +0.12 & +0.21 & +0.13 & +0.08 \\ \hline
Capped rate & +0.72 & +0.69 & +0.15 & +0.34 & +0.17 & +0.17 \\ \hline
Attenuated rate & +0.43 & +0.42 & +0.09 & +0.20 & +0.10 & +0.10 \\ \hline
10\% \acs{LOS} errors & +0.69 & +0.65 & +0.14 & +0.33 & +0.15 & +0.16 \\
\hline
\end{tabular}
\\[2pt]
\parbox{\columnwidth}{\scriptsize Offloading: SAATO + Joint-FP minus no offloading with per-tier FP. Joint-FP: SAATO + Joint-FP minus SAATO + per-tier FP. Geo: geometric mean; 5\%-ile: 5th percentile. All 95\% bootstrap intervals exclude zero. The first row uses new seeds, the others the drops of Table~\ref{tab:main}. In the two rate rows, the allocators still optimize the Shannon rate.}
\end{table}

Table~\ref{tab:robust} checks whether the gains persist under weaker assumptions. The results on 30 new drops match Table~\ref{tab:main}. With only large-scale channel knowledge or a 4~dB estimation error, all power-controlled schemes lose 0.7--0.9~Mbps in geometric-mean throughput (not tabulated), yet both gains remain positive. Attenuating the achievable rate lowers throughput by 35--43\% across schemes, while the relative gains remain 8\% and 2\% of the geometric mean of no offloading with per-tier \ac{FP}. The gains are also largely unchanged when the assumed \ac{LOS} state is inverted for 10\% of users, varying by at most 0.05~Mbps. Finally, at the 0.5~Mbps outage threshold, \ac{SAATO} with Joint-FP records no outage samples with instantaneous channel knowledge and at most 0.01\% with weaker knowledge. 

\section{Conclusion}
\label{sec:conclusion}
This paper developed a joint offloading and resource allocation framework for an \ac{ITNTN} in which a \ac{LEO} satellite beam and the terrestrial network operate over partially or fully overlapping frequency resources. The framework operates on two timescales: \ac{SAATO} selects the serving tier from filtered \ac{SINR} measurements of both candidate links, load-aware expected throughput and a hysteresis band, and Joint-FP allocates transmit power and \acp{PRB} with cross-tier interference-leakage terms, yielding a non-decreasing, convergent accepted utility sequence. The NP-hardness result concerns the sum-rate utility only. The framework was evaluated against an \ac{RSRP}-based baseline under different levels of spectral overlap and channel uncertainty. At full spectral overlap, the \ac{RSRP}-based rule left 44\% of user samples in outage, whereas \ac{SAATO} combined with Joint-FP left none below 0.5~Mbps given instantaneous channel knowledge, with no more than 0.05 tier switches per user per minute. Joint-FP alone reduced the outage rate of the \ac{RSRP}-based rule from 44.1\% to 26.5\%, and \ac{SAATO} with Joint-FP improved geometric-mean throughput over power control without offloading by 8\% and 2\% under half and full spectral overlap, respectively. These gains persisted under large-scale channel information, a 4~dB channel error and incorrect \ac{LOS} states. Future work will investigate multi-satellite operation, the \ac{UL}, delayed channel reports and adaptive spectral overlap.

\bibliographystyle{IEEEtran}
\bibliography{bib/ref}
 
\end{document}